\documentclass[11pt]{article}
\usepackage{amsfonts}
\usepackage{jfe}          
\usepackage{bm}
\usepackage{graphicx}
\usepackage{kbordermatrix}
\usepackage{array}
\usepackage{color}
\usepackage{mathrsfs}
\usepackage{tabularx,tabulary}
\usepackage{verbatim}
\usepackage{dsfont}
\usepackage{xcolor,colortbl}
\usepackage[sort]{natbib}
\usepackage[T1]{fontenc}
\usepackage{eurosym} 
\usepackage{lmodern}
\usepackage[normalem]{ulem}
\usepackage{bbm}
\usepackage{hyperref} 
\DeclareGraphicsExtensions{.jpg,.eps,.pdf} 
\usepackage{enumerate}
\usepackage{amsmath,amssymb,amsfonts,amsthm} 
\newtheorem{theorem}{Theorem}

\newtheorem{lemma}[theorem]{Lemma}

\theoremstyle{remark}

\definecolor{bleu}{cmyk}{1,0.4,0,0}

\newcommand{\E}{\mathbb{E}}

\newcommand{\R}{\mathbb{R}}

\newcolumntype{a}{>{\columncolor{Gray}}c}

\definecolor{Gray}{gray}{0.95}
\definecolor{White}{gray}{1.00}

\usepackage{xcolor}
\hypersetup{
citebordercolor=white,
filebordercolor=red,
linkbordercolor=blue
}
\hypersetup{
colorlinks,
citecolor=blue,
linkcolor=red,
urlcolor=blue}

\begin{document}


\title{Characteristics-driven returns in equilibrium}

\author{Guillaume Coqueret\thanks{EMLYON Business School, 23 avenue Guy de Collongue, 69130 Ecully, FRANCE. E-mail: coqueret@em-lyon.com}}
\maketitle
\begin{center}
\end{center}
\begin{abstract}
We reverse-engineer the equilibrium construction process of asset prices in order to obtain returns which depend on firm characteristics, possibly in a linear fashion. One key requirement is that agents must have demands that rely separately on firm characteristics and on the log-price of assets. Market clearing via exogenous (non-factor driven) supply, combined with linear demands in characteristics, yields the sought form. The coefficients in the resulting linear expressions are scaled net aggregate demands for characteristics, as well as their variations, and both can be jointly estimated via panel regressions. Conditions underpinning asset pricing anomalies are derived and underline the theoretical importance of the links between characteristics. Empirically, when the number of characteristics is small, the value and momentum anomalies are mostly driven by firm-specific fixed-effects, i.e., latent demands, which highlights the shortcomings of low-dimensional models.

\end{abstract}



\section{Introduction}

A central question in financial economics pertains to why assets experience different returns. There are mainly three streams of explanations to why that may be the case. First, assets may earn contrasting returns because they have idiosyncratic exposures to various sources of risk, or factors (as in, e.g., \cite{merton1973intertemporal}, \cite{ross1976arbitrage}, and \cite{fama1993common}). A second family of explanations relies on behavioural models, in which investor preferences or beliefs drive demand towards particular stocks, thereby generating heterogeneity in the cross-section of returns (\cite{barberis2003style}, \cite{barberis2015x}). Another angle related to these approaches is \textit{mispricing}, whereby agents overestimate or under-appreciate prices, often due to cognitive biases (see, e.g., \cite{lakonishok1994contrarian}, \cite{daniel1998investor}, \cite{hirshleifer2001investor} and \cite{stambaugh2017mispricing}). 
Finally, a third portion of the literature, which sometimes overlaps with the first two, argues that returns differ across assets simply because these assets \textit{are different}. It is their characteristics (their size, sector, balance sheet structure, past performance, riskiness of business lines, governance, etc.) which have an impact on their profitability (\cite{daniel1997evidence}). As such, these characteristics can be used as independent variables in predictive models, as in \cite{lynch2001portfolio} and \cite{gu2020empirical}, or they can be used to create more efficient risk factors (\cite{daniel2020cross}).

Technically, factors models benefit from the simplicity of parsimony because, in practice, a handful of variables are used to explain the entirety of cross-sectional differences in returns. Assets differ in their exposure to these variables: this is the source of heterogeneity in returns. However, one major issue is the definition or identification of the factors. When they are latent (as in \cite{chamberlain1983funds}, \cite{connor1993test}, \cite{bai2002determining}, or, more recently, in \cite{kelly2019characteristics}, and \cite{lettau2020estimating,lettau2020factors} to cite but a few), they are often hard to understand - though recent advances propose enlightening interpretations (see \cite{clarke2021level}). When they are explicit and based on prior empirical conclusions (e.g., SMB or HML in \cite{fama1993common} or WML from \cite{jegadeesh1993returns}), they remain somewhat arbitrary.\footnote{It is nonetheless a very interesting exercise to theoretically justify factors \textit{a posteriori}, akin to transparent HARKing (\cite{hollenbeck2017harking}). This has generated an insightful literature, and we refer for instance to the following common anomalies: \vspace{-1mm}\begin{itemize}
\setlength\itemsep{-0.2em}
\item \textbf{size}: \cite{berk1995critique};
\item \textbf{value}: \cite{zhang2005value}, \cite{lettau2007long};
\item \textbf{size} and \textbf{value}: \cite{gomes2003equilibrium}, \cite{campbell2004bad}, \cite{arnott2015can};
\item \textbf{momentum}: \cite{hong1999unified}, \cite{grinblatt2005prospect}, \cite{biais2010equilibrium}, \cite{vayanos2013institutional}, \cite{choi2014momentum}.
\end{itemize} \vspace{-5mm}
}

On the other hand, firm characteristics are directly observable, though subtleties in measurement can lead to diverging results.\footnote{This is well documented for instance for the value factor (\cite{asness2013devil} and \cite{hasler2021value}), and in sustainable investing (\cite{dimson2020divergent} and \cite{avramov2020investment}) - not to mention firm-specific sentiment.} Several studies have sought to single out the characteristics which drive the cross-section of stock returns (\cite{green2017characteristics}, \cite{penman2018framework}, \cite{freyberger2020dissecting}, and \cite{han2020firm}). In addition, some contributions have compared the efficiency of characteristics-based versus factor-based models in asset pricing.\footnote{It is interesting to note that a recent trend in the literature expects to get the best of both worlds by resorting to characteristics \textit{within} factor models, their alphas and their loadings: see \cite{cederburg2015asset}, \cite{cosemans2016estimating}, \cite{dittmar2017firm}, \cite{kelly2019characteristics}, \cite{connor2021dynamic}, \cite{ge2021dynamic}, \cite{kim2021arbitrage,kim2021characteristic} and \cite{windmuller2021firm}.} For instance, \cite{hou2011factors}, \cite{goyal2018cross}, \cite{chordia2019cross}, \cite{fama2020comparing}, \cite{raponi2020testing} and \cite{chib2021slope}) all report that the former have a better explanatory power over stock returns. \cite{jacobs2021factor} provide some historical perspectives on the differences between the two approaches. On the practitioners' side, \cite{brightman2021surprise} find that characteristics generate more accurate predictions and, thus, are more likely to deliver out-of-sample alpha.

The purpose of the present article is to provide a theoretical grounding for models that use firm characteristics as predictors of future returns. The construction of a partial equilibrium\footnote{Recently, general equilibrium models based on characteristics have emerged. We refer for example to \cite{alti2019dynamic}, \cite{betermier2020supply}, \cite{betermier2021portfolios}. In \cite{buss2021dynamics}, demand depends essentially on one latent variable.} is decomposed and the demand of agents is tailored to yield a special form for asset (log-)returns. The sought linear expression is obtained under particular assumptions. Notably, agents are required to build their allocations (net demands) as linear functions of characteristics plus a multiple of the log-price of assets. We show that these allocations can be decomposed as linear combinations of characteristics if agents believe that expected returns are linear functions of characteristics. Loosely speaking, the model then becomes self-fulfilling because the agents' assumption materializes in equilibrium. Nonetheless, one important suggestion from the model is that returns should also depend on characteristics' variations, and not only on their levels. 

The derivation of equilibrium returns provides a novel interpretation for the loadings of predictive panels relying on firm characteristics. All estimated coefficients pertain to \textit{scaled} demands in characteristics, and changes thereof. Therefore, the model is able to extract the relative demand in each characteristic - all without any data on flows or holdings. The outcome gives two types of insights. The first is the sign of the demand, e.g., when investors are on average more ``value'' than ``growth'', so to speak. The second is the relative scale of demands across factors, i.e., when agents for instance seek size more than momentum, or vice-versa.

The expressions for individual returns can be aggregated into sorted portfolio returns. This reveals the importance of characteristics' interactions in the decomposition of anomalies. The prevalence of the latter for asset pricing is also documented in \cite{ross2021characteristic} and \cite{duan2021factorization}. Three terms emerge in the decomposition of long-short portfolio returns: the loading of the characteristic used for the sorting, the average fixed effect of the portfolio, and the covariance terms. Our empirical results indicate that when the number of characteristics is small (three or six), the fixed effect term often dominates the loading and the interaction components compensate this asymmetry. This is in line with the findings of \cite{koijen2019demand}, who also document the prominent impact of latent demands. Nevertheless, the magnitudes of the terms are strongly dependent on the size of the rolling windows that are used for the estimations. As sizes increase, the prevalence of fixed effects decreases.

The remainder of the article is structured as follows. In Section \ref{sec:model}, we present the model and some of its theoretical properties. In Section \ref{sec:ano}, we investigate the implications for asset pricing anomalies. Section \ref{sec:esti} is dedicated to the empirical study. Finally, Section \ref{sec:conc} concludes. Three proofs and additional material are located in the Appendix.

\section{The model}
\label{sec:model}

\subsection{Discussion on partial economic equilibria}
In financial economics, the simplest market clearing equation is the following:
\begin{equation}
\bm{d}_t(\bm{p}_t) = \bm{s}_t(\bm{p}_t),
\label{eq:clear0}
\end{equation}
where the vectors of aggregate demands $\bm{d}_t$ and supplies $\bm{s}_t$ depend on the vector of prices $\bm{p}_t$. To ease the calculations, it is customary to model the demand side only. One rationale for this choice is that researchers prefer to investigate the impact of the demand side on asset prices. Often, investors are separated into heterogeneous groups. One group will craft its allocation decisions based on preferences as well as on some information set, including the price of the asset, while the other group is considered as market maker (liquidity provider, the right side of \eqref{eq:clear0}) and trades independently for the price level. The equation then becomes
\begin{equation}
\bm{d}_t(\bm{p}_t) = \bm{s}_t \quad \Longrightarrow \quad \bm{p}_t = \bm{d}_t^{-1}(\bm{s}_t),
\label{eq:clear1}
\end{equation}
where the implication only holds when the inverse (multivariate) mapping $\bm{d}_t^{-1}$ is well-defined. Going into further detail, the total demand function $\bm{d}_t$ can be broken down if we consider heterogeneity in the demand of agents, in which case, 
\begin{equation}
\sum_{i=1}^IA_{t,i}\bm{w}_{t,i}(\bm{p}_t) = \bm{s}_t,
\label{eq:clear2}
\end{equation}
where $A_{t,i}$ is the time-$t$ wealth of agent $i$ that is invested on the market and $\bm{w}_{t,i}$ is the corresponding relative buy or sell quantities (strictly speaking they are not necessarily portfolio compositions and we discuss this nuance later on). One favorable case is when this demand form can be factorized into:
\begin{equation}
\left(\sum_{i=1}^IA_{t,i}\bm{w}_{t,i}\right)(\bm{p}_t) = \bm{s}_t  \quad \Longrightarrow \quad\bm{p}_t =  \left(\sum_{i=1}^IA_{t,i}\bm{w}_{t,i}\right)^{-1}(\bm{s}_t  ),
\label{eq:clear3}
\end{equation}
where, again, the implication only holds if the inverse makes sense. The factorization can occur when $\bm{w}_{t,i}$ is separable, i.e., $\bm{w}_{t,i}=w_i \times \bm{w}_t(\bm{p}_t)$, or when the price-driven part is linear, that is, when $\bm{w}_{t,i}=a_{t,i,n}+b_{t,i}\bm{p}_t$. Under reasonable assumptions, the $b_{t,i}$ in the latter form is supposed to be negative, because demand usually decreases with price.\footnote{There is an ongoing debate on whether demand curves for stocks slope down. Empirical analyses from \cite{shleifer1986demand}, \cite{kaul2000demand}, \cite{petajisto2009demand} and \cite{buss2021dynamics} support this conjecture, but those in \cite{cha2001market} and \cite{jain2019demand} do not. \cite{wurgler2002does} conclude that it depends on whether or not the stock has substitutes on the market. In \cite{koijen2019demand}, the uniqueness of prices requires that demands be strictly downward sloping for all investors.} Such linear forms can be obtained via mean-variance preferences (see, e.g., \cite{admati1985noisy} and \cite{kacperczyk2019investor}).

The main issues with most theoretical models is that they yield \textit{prices} and not \textit{returns}. If we want to obtain returns from Equation \eqref{eq:clear3}, we must tackle the following expressions (logarithmic versus arithmetic returns):
\begin{align}
\bm{r}_{t+1}&=\log \left(\text{diag}\left( \left(\sum_{i=1}^IA_{t,i}\bm{w}_{t,i}\right)^{-1}(\bm{s}_t  ) \right)^{-1}  \left(\sum_{i=1}^IA_{t+1,i}\bm{w}_{t+1,i}\right)^{-1}(\bm{s}_{t+1}  )\right), \quad \text{or} \\
 \bm{r}_{t+1}&  =\text{diag}\left( \left(\sum_{i=1}^IA_{t,i}\bm{w}_{t,i}\right)^{-1}(\bm{s}_t  ) \right)^{-1}  \left(\sum_{i=1}^IA_{t+1,i}\bm{w}_{t+1,i}\right)^{-1}(\bm{s}_{t+1}  )-1,
\label{eq:ret0}
\end{align}
where diag$(\bm{v})$ fills a diagonal matrix with the values of vector $\bm{v}$. The two expressions above are impractical to work with in all generality. It is therefore imperative to impose a strong structure on the agent demands $\bm{w}_{t,i}$ to obtain tractable formulae for returns. This is the purpose of the next section. Closed-form expressions are not necessary for empirical applications as long as prices or returns can easily be evaluated numerically (as is done in \cite{koijen2019demand}). Nonetheless, they often offer insightful interpretations.

\subsection{Characteristics-based demands and returns}

In this subsection and henceforth, we  assume agents allocate according to firms' characteristics. By construction, this generates a characteristics-based structure for assets' log-returns. 

Time is discrete and denoted with $t$. Investors (or agents) on the market are indexed with $i=1\dots,I$ and they trade between $N>1$ assets, which are indexed with $n=1,\dots,N$. We write $p_{t,n}$ for the time-$t$ price of asset $n$. In addition to their prices, all assets are characterized by exactly $K$ indicators, $c_{t,n}^{(k)}$, for $k=1,\dots,K$. These indicators are publicly disclosed and available to all agents on the market. Common examples for equities include market capitalization (firm size), accounting and valuation ratios, risk measures (volatility) and past performance (stock momentum).\footnote{It could be debated whether alternative metrics, like stock-specific sentiment or ESG-related data fall into this category. This discussion is outside the scope of this paper.} Examples for bonds encompass durations or credit ratings. In the present paper, we will restrict our analysis and examples to the case of stocks, but our theoretical results hold for any asset class, as long as the price is determined by the market clearing mechanism mentioned above.

The $c_{t,n}^{(k)}$ need not be raw values, but can represent synthetic scores which are scaled in the cross-section of stocks, as is now commonplace in the literature on characteristics-based factor models (e.g., \cite{koijen2019demand}, \cite{kelly2019characteristics} and \cite{freyberger2020dissecting}). We write $\bm{c}_{t,n}$ for the time-$t$ $K$-dimensional vector of characteristics of asset $n$.

One central hypothesis of the model is that the weights (or demands) $\bm{w}_t$ are unconstrained and can be negative. For instance, this can correspond to the case where market clearing operates on \textit{net} demands. Markets and agents would be assumed to be mature so that, at each time step, the latter \textit{adjust} their portfolio by fine-tuning pre-existing positions. Following \cite{koijen2019demand}, we assume that these demands are driven by agent's preferences towards assets' characteristics. This is a rather weak assumption because our definition of characteristics is large and it makes sense that agents form their allocation based on some observable criteria. \cite{koijen2019demand} show that characteristics-based demands can be viewed as optimal if characteristics are informative for the evaluation of the first two moments of expected returns. Several studies document the preference of certain investors for particular characteristics (\cite{froot2008style}, \cite{kumar2009dynamic}, \cite{cronqvist2015value}, \cite{betermier2017value}, \cite{koijen2020investors} and \cite{balasubramaniam2021owns}), so that optimality is not necessarily an imperative requirement. 

In addition to characteristics-driven investors, there exist \textit{external} agents who trade purely orthogonally to these attributes and act as market makers in our model. Consequently, they provide a net supply for each stock, which we write $s_{t,n}$. We will not further mention these \textit{external} agents, except via this exogenous supply which they provide.

Because we reason in terms of \textit{net} demand, we cannot resort to an exponential function, as in \cite{koijen2019demand}, because net demands can be negative, e.g., when an agent wants to reduce a position in an asset, or sell it short. Instead, we work with the general form
\begin{equation}
w_{t,i,n}=a_{t,i,n}+ b_{t,i}^{(0)}f(p_{t,n})+ g_{t,i}(\bm{c}_{t-1,n}),
\label{eq:demand0}
\end{equation}
for the time-$t$ demand of agent $i$ in asset $n$. We will study particular shapes for the functions $f$ and $g_{t,i}$ subsequently.
The above demand is expressed as a percentage of investor $i$'s wealth, i.e., it can be considered as a portfolio composition, even though we do not impose that it sums to one across all $N$ firms. 

The rightmost part of \eqref{eq:demand0} implies that the factor-driven agents construct their portfolios based on indicators which they observe or receive at time $t-1$. In practice, it is not uncommon that investors wait for quarterly updates in accounting disclosures before they rebalance their portfolios. This convention does not affect most of the results in the paper. Simply put, if characteristics' time index is lagged ($\bm{c}_{t-1}$), they are predictors. If it is synchronous ($\bm{c}_t$), then the model will explain returns but not predict them. Portfolio policies that are linear in firm characteristics are widespread in the literature (see, e.g., \cite{brandt2009parametric}, \cite{hjalmarsson2012characteristic}, and \cite{ammann2016characteristics}) and they will constitute an important special case subsequently.

The separation of the price $p_{t,n}$ from the other characteristics $\bm{c}_{t-1,n}$ in \eqref{eq:demand0} is a crucial technical requirement. In any equilibrium-based asset pricing model, the demand function of at least some agents need to be expressed in terms of the price of the asset, which becomes the unknown in a market clearing equation. Solving for this unknown yields the equilibrium price. In \cite{koijen2019demand}, this is implicitly done via market equity, which is factorized into its price versus number of shares components. 

The constant $a_{t,i,n}$ in Equation \eqref{eq:demand0} tunes the demand of agent $i$ towards asset $n$, regardless of the asset's attributes. It could for instance be driven by macro-economics factors, or private information. Technically, it could be made stock-independent, so that $a_{t,i}$ would evaluate the global equity exposure of the agent that is independent from the characteristics. From an estimation standpoint, the $a_{t,i,n}$ will allow for an interesting interpretation which we mention later on. The central component $ b_{t,i}^{(0)}f(p_{t,n})$ is an important technical artefact that is used to build the equilibrium price. Let us simply assume for now that the function $f$ is monotonic. 

If we denote with $A_{t,i}$ each agent's wealth that is subject to trading at time $t$, then market clearing imposes that for each asset, total net demand matches total net supply, i.e.,
\begin{equation}
\sum_{i=1}^I A_{t,i} w_{t,i,n} = s_{t,n}.
\label{eq:clear6}
\end{equation}

The most important assumption of the model is the separation, in the demand, between the \textit{log}-price and the other characteristics, which are unrelated to the former.\footnote{Obviously, market capitalization or valuation ratios incorporate the price of the asset. But we assume for analytical tractability that the scores $\bm{c}_{t,n}$ are unrelated to asset prices. If characteristics are synthetic indicators which are scaled in the cross-section of stocks, this hypothesis is not too far-fetched.} This implies
\begin{equation}
\sum_{i=1}^I A_{t,i}\left( a_{t,i,n}+ b_{t,i}^{(0)}\log(p_{t,n}) + g_{t,i}(\bm{c}_{t-1,n})\right)= s_{t,n},
\label{eq:dem0}
\end{equation}
i.e.,
\begin{equation}
\log(p_{t,n})=\frac{\overbrace{\sum_{i=1}^I A_{t,i} \left(a_{t,i,n}+ g_{t,i}(\bm{c}_{t-1,n}) \right)}^{\text{total non-price related demand}}-\overbrace{s_{t,n}}^{\text{supply}}}{-\underbrace{\sum_{i=1}^IA_{t,i}b_{t,i}^{(0)}}_{\text{agg. demand for log-price}}} .
\end{equation}
Intuitively, it seems reasonable to assume that the denominator $-\sum_{i=1}^IA_{t,i}b_{t,i}^{(0)}$ is positive, because we expect prices to decrease with supply. This amounts to posit that the aggregate demand for log-prices is negative, which is a reasonable postulate. In \cite{koijen2019demand}, demand slopes are negative for all investors, while we only require the aggregate slope to be negative.
This gives a simple formula for logarithmic returns:
\begin{align}
r_{t+1,n}&=\log\left(\frac{p_{t+1,n}}{p_{t,n}}\right)  \nonumber \\
&=\frac{\sum_{i=1}^I A_{t+1,i} \left(a_{t+1,i,n}+g_{t+1,i}(\bm{c}_{t,n})\right)-s_{t+1,n}}{\kappa_{t+1}} -\frac{\sum_{i=1}^I A_{t,i} \left(a_{t,i,n}+g_{t,i}(\bm{c}_{t-1,n})\right)-s_{t,n}}{\kappa_t}  \nonumber\\
&=\sum_{i=1}^I B_{t+1,i} \left(a_{t+1,i,n}+g_{t+1,i}(\bm{c}_{t,n})\right)- \sum_{i=1}^I B_{t,i} \left(a_{t,i,n}+g_{t,i}(\bm{c}_{t-1,n})\right)+   \frac{s_{t,n}}{\kappa_{t}}-\frac{s_{t+1,n}}{\kappa_{t+1}}  \\
&\small =\underbrace{\underbrace{\sum_{i=1}^I (B_{t+1,i} a_{t+1,i,n}-B_{t,i}a_{t,i,n})}_{\substack{\text{change in scaled} \\ \text{ non-characteristic demand}}} + \underbrace{\sum_{i=1}^I (B_{t+1,i} g_{t+1,i}(\bm{c}_{t,n})-B_{t,i}g_{t,i}(\bm{c}_{t-1,n}))}_{\substack{\text{change in scaled} \\ \text{ pure characteristic demand}}}}_{{g}^*(\bm{c}_{t,n}, \bm{c}_{t-1,n})}+ \underbrace{ \underbrace{ \frac{s_{t,n}}{\kappa_{t}}-\frac{s_{t+1,n}}{\kappa_{t+1}} }_{\text{supply shock}}}_{e_{t+1,n}} \label{eq:center}
\end{align}
where $\kappa_t=-\sum_{i=1}^IA_{t,i}b_{t,i}^{(0)}>0$ is minus the aggregate demand for the log price and $B_{t,i}=A_{t,i}/\kappa_t$ are the scaled wealths. On purpose, we simplified the expression in the last line (at the root of the horizontal brackets) by using the same notations as in Equations (1) and (2) from \cite{gu2020empirical}. This underlines that the result can be used as theoretical justification to rely on model-agnostic machine learning techniques (based on characteristics) when modelling the cross-section of asset returns.

Before we further specify the demand function, it is useful to discuss two weaknesses in the above formulation. The first one is that the decomposition only works for logarithmic returns. While they are close to their arithmetic counterparts, some difference may arise in the cross-section of mean returns, especially if assets' volatilities differ. This comes from a simple application of the Taylor series of the mapping $x\mapsto \log(1+x)$ (see \cite{hudson2015calculating}). The second drawback of the above expression is that it does not incorporate dividends: returns are hence \textit{price} returns and not \textit{total} returns. The latter are therefore out of the scope of our analysis.

\subsection{The case of linear demands}

An appealing special case of the demand component $g_{t,i}$ is the linear combination of characteristics. This is for instance exploited in \cite{brandt2009parametric}, or in \cite{koijen2019demand} when the linear form is exponentiated. The expression is simply:
\begin{equation}
g_{t,i}(\bm{c}_{t-1,n})=a_{t,i,n}+\sum_{k=1}^K  b_{t,i}^{(k)} c_{t-1,n}^{(k) },
\label{eq:lindem}
\end{equation}
where the constants $ b_{t,i}^{(k)}$ determine the sign and appetite intensity of agent $i$ for characteristic $k$. In \cite{koijen2019demand}, such linear forms are obtained when agents believe in a single factor model in which the first two moments of returns are estimated through firm characteristics. In Lemma \ref{lem:6} below, we show that linear demands can be obtained via another theoretical route. We recall that, for agent $i$, the expression for a mean-variance optimal portfolio has the form 
$$\bm{w}^*_{t,i,n}= \gamma_{t,i}^{-1} \mathbb{V}_{t,i}[\bm{r}_{t+1}]^{-1}(\bar{\bm{r}}_t+ \delta_{t,i} \bm{1}),$$ 
where $\bar{\bm{r}}_t $ is the time-$t$ vector of expected returns, $\gamma_{t,i} $ is the time-$t$ risk aversion of agent $i$ , and $ \delta_{t,i}$ is a scalar chosen to satisfy the budget constraint. Under particular beliefs, this expression can be factorized in a particular form, as stated below.

\begin{lemma}
\label{lem:6}
If agent $i$ believes that returns are driven by 
\begin{equation}
\bm{r}_{t+1}=\bm{C}_t\bm{\beta}_{t+1,i}+ \bm{e}_{t+1},
\label{eq:ass}
\end{equation}
then the optimal budget-constrained mean-variance portfolio weight for asset $n$ can be written as
\begin{align}
\bm{w}^*_{t,i,n}= f_{i,n,1}+\sum_{k=1}^Kc_{t,n}^{(k)}\times f_{i,n,2}, \label{eq:f0}
\end{align}
where $f_{i,n,1}:=f_{i,n,1}(\bm{C}_t,\hat{\bm{\beta}}_{t,i},\hat{\bm{\Sigma}}_{\bm{\beta},i}, \hat{\bm{\sigma}}_{e,i}^2)$ and $f_{i,n,2}:=f_{i,n,2}(\bm{C}_t,\hat{\bm{\beta}}_{t,i},\hat{\bm{\Sigma}}_{\bm{\beta},i}, \hat{\bm{\sigma}}_{e,i}^2)$ are scalars that depend on the data $\bm{C}_t$, as well on agent $i$'s estimations for the terms in Equation \eqref{eq:ass}.
\end{lemma}

The proof of the lemma is located in Appendix \ref{sec:lindem}. Of course, strictly speaking, the weights are not purely linear in the characteristics, because the latter are present in the definition of the functions $f_{n,1}$ and $f_{n,2}$.  
The only difference between \eqref{eq:f0} and \eqref{eq:lindem} is the time shift in the characteristics from $c_{t,n}^{(k)}$ to $c_{t-1,n}^{(k)}$. To take into account time lags in the diffusion of the information, we henceforth stick with the latter.  

Formally, it is always possible to express agent demands as a linear function of characteristics, as long as an error term is allowed, which is how \cite{koijen2019demand} proceed. The equation $\bm{w}=\bm{Cb}$ has exactly one solution if $\bm{C}$ is square ($N=K$) and non-singular. It has an infinite number of solutions if $K>N$, and it has no solution if $K<N$, in which case $\bm{w}$ equals $\bm{Cb}$ plus an additional error term. In the specification \eqref{eq:lindem}, the constant $a_{t,i,n}$ can be assimilated to this error term, as it captures the demand that is not driven by the characteristics.

The demand form \eqref{eq:lindem} allows to change the notation and include the $a_{t,i,n}$ terms in the sum, so that the linearized form of Equation \eqref{eq:center} now reads 
\begin{equation}
r_{t+1,n} =\sum_{i=1}^I\left(B_{t+1,i}a_{t+1,i,n}-B_{t,i}a_{t,i,n}+ \sum_{k=1}^{K} \left( B_{t+1,i}  b_{t+1,i}^{(k)} c_{t,n}^{(k) }   -  B_{t,i}  b_{t,i}^{(k)} c_{t-1,n}^{(k) } \right)\right) + \varepsilon_{t+1,n},
\end{equation}
where
\begin{equation}
\varepsilon_{t+1,n}=\frac{s_{t,n}}{\eta_{t}}-\frac{s_{t+1,n}}{\eta_{t+1}} 
\label{eq:charfree}
\end{equation} 
is the innovation from the supply-side. We can then swap the two sums (in $i$ and $k$) in the central term and, for a given $k$, we can decompose the central shift in two ways, depending one the factors we put forward: \vspace{-6mm}

\begin{align} 
&\sum_{i=1}^I \left(c_{t,n}^{(k) } B_{t+1,i} b_{t+1,i}^{(k)}-c_{t-1,n}^{(k) }B_{t,i} b_{t,i}^{(k)}\right) \\ =& \ c_{t,n}^{(k)} \underbrace{\sum_{i=1}^I \left( B_{t+1,i}b_{t+1,i}^{(k)}- B_{t,i}b_{t,i}^{(k)}\right)}_{\beta^{(k)}_{t+1}=\, \text{change in scaled agg. demand}} + \underbrace{\left(c_{t,n}^{(k)} -c_{t-1,n}^{(k)}\right)}_{\text{past change in char.}} \underbrace{\sum_{i=1}^I B_{t,i}b_{t,i}^{(k)}}_{\eta^{(k)}_t= \, \text{past demand}}& (\text{first identity}) \label{eq:def7} \\
=&  \ \eta_{t+1}^{(k)}(c_{t,n}^{(k)}-c_{t-1,n}^{(k)}) + c_{t-1,n}^{(k)}\beta_t^{(k)}.&(\text{second identity}) 
\end{align}

In the above expressions, we use the term ``\textit{demand}'' slightly improperly. $\eta_t^{(k)}$ and $\beta_t^{(k)}$ defined in the brackets are in fact the purely characteristics driven \textit{components} of the demand. The factor $\beta_t^{(k)}$ can be viewed as the aggregate willingness to be exposed to the characteristics, while $\eta_t^{(k)}$ signals willingness to be exposed to \textit{variations} in characteristics. However, we will henceforth resort to this abuse of language and notation and refer to these terms as demands.

Following Section 9-D in \cite{fama1976foundations}, \cite{fama2020comparing} argue that if returns are linear functions of past values of firm characteristics, then the corresponding loadings (or slopes) are \textit{returns} of long-short portfolios. We underline that this statement only holds if loadings are both time-dependent and are estimated via least square minimization.\footnote{Indeed, the expression $\bm{r}=\bm{Xb}+\bm{e}$ has OLS coefficients $\hat{\bm{b}}=(\bm{X}'\bm{X})^{-1}\bm{X}'\bm{r}$, which is a linear combination of returns, and hence, a portfolio (see also \cite{kirby2020firm}). Interestingly, \cite{stevens1998inverse} proves a reverse identity in which optimal portfolios can be obtained via regressions, though in this case the dependent variables are other assets' returns - see \cite{goto2015improving} and \cite{deguest2018reinterpretation} for extensions of this approach. The links between regressions and portfolios also include the intuitive interpretations of Theorem 1 in \cite{britten1999sampling} and in \cite{leung2021statistical}.} 

In the two identities, the slope associated to a particular characteristic is the change in scaled demand for this characteristic. Heuristically, scaled demands (\textit{resp.}, changes in demands) can be viewed as portfolio weights (\textit{resp.}, changes in portfolio weights). But importantly, equilibrium returns are also driven by \textit{variations} in the characteristics as well. This seems to indicate that momentum in firm attributes should be more investigated in the literature. Several contributions tackle this topic (e.g., \cite{ohlson1992changes}, \cite{chen2003characteristics}, \cite{novy2015fundamentally}, \cite{blank2020corporate}, and, to a certain extent, \cite{gabaix2021search} when the characteristic is the expected dividend). Given the decompositions outlined above, we express the logarithmic returns in a simple form. 

\begin{lemma}
\label{lem:1}
We assume that market clearing is defined in \eqref{eq:clear6} and demands satisfy \eqref{eq:demand0} and \eqref{eq:lindem}. In partial equilibrium, it holds that, \vspace{-2mm}
\begin{align}
r_{t+1,n}&=\alpha_{t+1,n} + \sum_{k=1}^{K} \left(\beta_{t+1}^{(k)}c_{t,n}^{(k)} + \eta_t^{(k)}\Delta c_{t,n}^{(k)} \right) + \varepsilon_{t+1,n}, \quad \text{or} \label{eq:simple} \\ \vspace{-2mm}
r_{t+1,n}&=\alpha_{t+1,n} + \sum_{k=1}^{K} \left(\beta_{t}^{(k)}c_{t-1,n}^{(k)} + \eta_{t+1}^{(k)}\Delta c_{t,n}^{(k)} \right) + \varepsilon_{t+1,n}, \label{eq:simpl2} \vspace{-2mm}
\end{align}
where $\Delta c_{t,n}^{(k)}=c_{t,n}^{(k)} -c_{t-1,n}^{(k)}$ is the local change in the characteristic, $\beta_{t}^{(k)}$ is the change in scaled aggregate demand for characteristic $k$, and $\eta_{t}^{(k)}$ is the scaled demand for characteristic $k$ defined in Equation \eqref{eq:def7}. Innovations terms $\varepsilon_{t+1,n}$ come from the supply side and are given in \eqref{eq:charfree}. Finally, the stock-specific constant is the change in aggregate scaled demand that is not driven by characteristics: $\alpha_{t+1,n}=\sum_{i=1}^I(B_{t+1,i}a_{t+1,i,n}-B_{t,i}a_{t,i,n})$.
\end{lemma}

We list a few comments on this result below. \vspace{-2mm}
\begin{itemize}
\setlength{\itemsep}{-3pt}
\item In the main equation of the lemma, the time index for the characteristics could be shifted from $t$ to $t+1$ and from $t-1$ to $t$. This corresponds to the case when agents form their portfolios based on synchronous data in Equation \eqref{eq:demand0}.
\item The constant $\alpha_{t+1,n}$ is very convenient for estimation purposes. If the unknowns $\beta^{(k)}$ and $\eta^{(k)}$ in equations in the lemma were to be estimated via a panel approach, $\alpha_{t+1,n}$ could be interpreted as a \textbf{fixed}, \textbf{random}, or \textbf{between effect}, depending on the modelling assumptions. Using the terminology of \cite{koijen2019demand}, it is the average latent demand of asset $n$.
\item One implication of the Lemma is that the time-$t$ conditional expectation of the returns is given by \vspace{-4mm}
\begin{align}
\mathbb{E}_t[r_{t+1,n}] &= \sum_{k=1}^{K}\left( \eta_t^{(k)}\Delta c_{t,n}^{(k)} +c_{t,n}^{(k)} \mathbb{E}_t\left[\beta_{t+1}^{(k)}  \right] \right)+  \mathbb{E}_t[\alpha_{t+1,n}+\varepsilon_{t+1,n}]
\label{eq:condexp} \\
&=\sum_{k=1}^{K}\left(\beta_{t}^{(k)}c_{t-1,n}^{(k)} + \Delta c_{t,n}^{(k)} \mathbb{E}_t\left[\eta_{t+1}^{(k)} \right] \right)+  \mathbb{E}_t[\alpha_{t+1,n}+\varepsilon_{t+1,n}].
\end{align}
In short, and quite naturally, the conditional average returns depend on expected changes in aggregate demand and aggregate supply, plus some term that reflects aggregate preferences that are orthogonal to characteristics. These expressions hold in all generality under the assumptions stated above. One important caveat is that the model is contingent on the choice of the characteristics. It is of course impossible to list all indicators monitored and considered by all agents around the world. Implicitly, indicators that are omitted are integrated in the right part of the equations, that is, in the asset-specific term or in the supply side of the model.
\item Because of the simple linear form, the \textit{pricing power} of characteristics and their variations is directly linked to the magnitude of the corresponding demand. A characteristic $c$ with large absolute demand $|\eta|$ will have a sizeable impact via $\Delta c$, while a change $\Delta c$ with an important absolute demand $|\beta|$ will strongly drive returns through $c$. Reversely, characteristics with negligible demands will only play a marginal role. 
\item The relationship in the lemma is \textit{predictive} because of the time lag between characteristics $\bm{c}_{t}$ and returns $\bm{r}_{t+1}$. If agents trade rapidly based on $\bm{c}_t$ in Equation \eqref{eq:demand0}, the lag vanishes and the characteristics can only \textit{explain} returns but not forecast them.
\item In a sense, Lemma \ref{lem:1} is self-fulfilling. If we posit that all non-supply side agents believe that returns are linear mappings of characteristics as in Lemma \ref{lem:6}, then, loosely speaking, their assumption materializes in Lemma \ref{lem:1}, if we omit the $\Delta$ terms.
\end{itemize}


\subsection{A word on log-price demands}

Logarithmic forms such as the one in the left side of Equation \eqref{eq:dem0} are rare in the literature because economists usually prefer to avoid negative demands.\footnote{Though, in all generality, linear forms such as those in \cite{admati1985noisy} and \cite{kacperczyk2019investor} can also be subject to negative demands.} In a financial framework in which agents hold shares and adjust their portfolios, negative demands reflect a decision to reduce positions in particular assets. Likewise, if agents are allowed to sell stocks short (via negative portfolio weights), we consider that they are expressing a negative demand.

If the simple links between returns and characteristics in Lemma \ref{lem:1} have some empirical validity, this would imply that agents' demands can be modelled with a logarithmic form.\footnote{Conversely, portfolio holdings from retail or institutional investors could corroborate or invalidate this form, thereby confirming or contradicting the result of the lemma. This is out of the scope of the paper.}

As a function of the price, the demand is simply written 
\begin{equation}
w=a-b\times \log(p), 
\label{eq:dem}
\end{equation}
where $a>0$ is the appeal of the the object (the stock) and $b>0$ is the slope. The corresponding elasticity is equal to $b(a-b\log(p))^{-1}$. Notably, it increases to infinity when the price shrinks to zero. While this could be an issue, we recall that most studies in asset pricing exclude penny stocks from empirical analyses.

\begin{figure}[!h]
\begin{center}
\includegraphics[width=12cm]{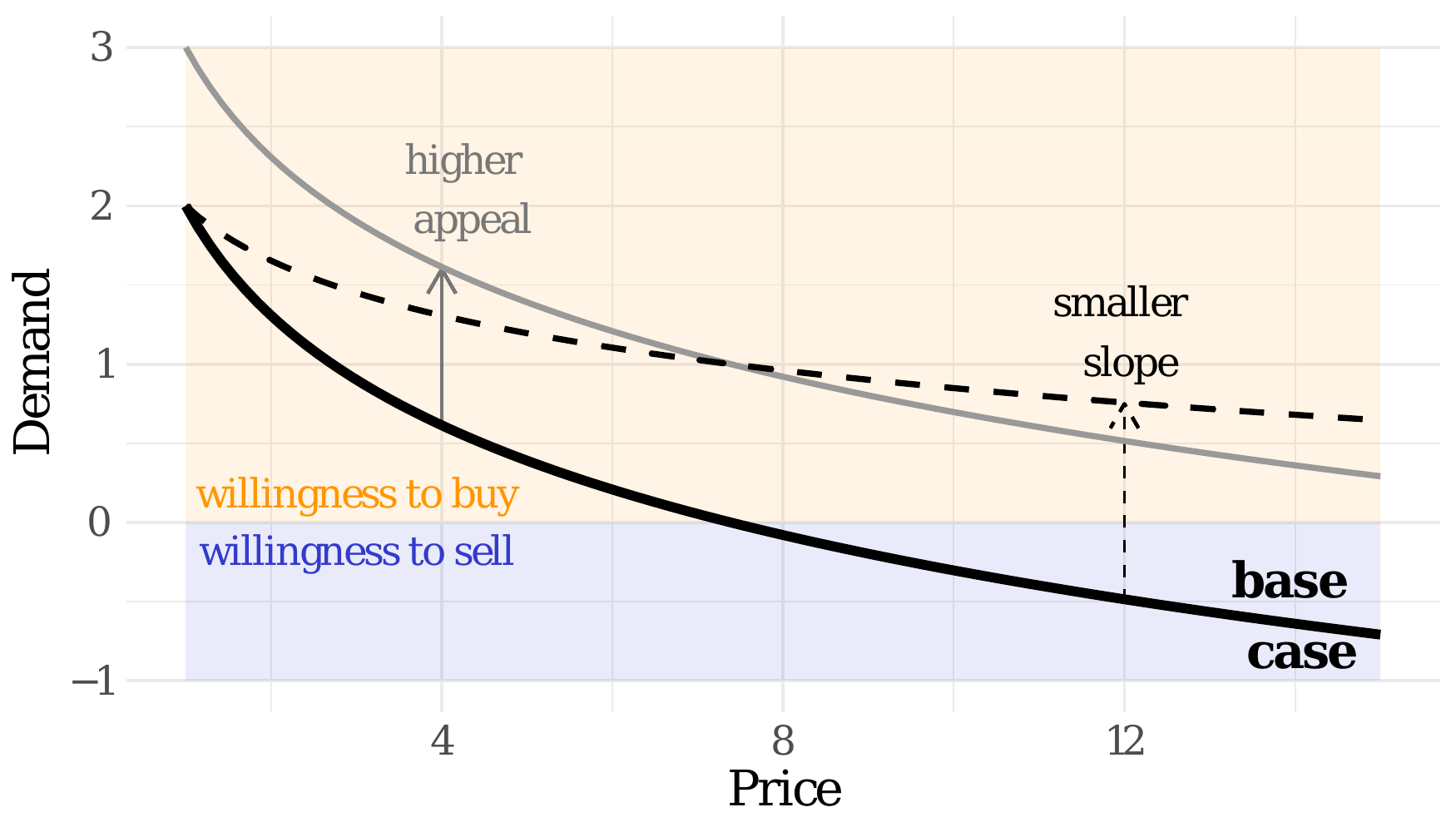}\vspace{-3mm}
\caption{\textbf{Demand curve}. \small This diagram shows the demand as a function of the price. The functions are equal to $\text{demand} = \text{appeal} - \text{slope} \times \text{log(price)}$. For the base case, the appeal is equal to two and the slope to one. The grey line has an appeal of three, while the dashed curve has a slope of $1/2$.}
\label{fig:demand}
\end{center}
\end{figure}

In our model, the appeal is solely driven by the characteristics (plus a constant term). The slope is purely investor-specific. The relationship is illustrated in Figure \ref{fig:demand}. The demand can be increased by higher appeal, or lower slope, or both.

This form is very opportune for aggregation, which is one property we have largely exploited in the previous subsection. If there are $I$ agents, each with demand $w_i=a_i-b_i\log(p)$, the latter can be summed to obtain \eqref{eq:dem}, where $w$, $a$, and $b$ are the sums of the respective individual values.
In addition, the form \eqref{eq:dem}, when equated to some exogenous supply $s$ yields $p=\exp \left(\frac{a-s}{b}\right)$ which has intuitive interpretations.


\section{Asset pricing anomalies}

\label{sec:ano}

\subsection{Sorted portfolios}

A cornerstone result in financial economics is the capital asset pricing model (the CAPM, for which we refer to \cite{perold2004capital} for a historical perspective). The CAPM states that individual stock returns must solely be driven by their exposure to the aggregate market return. Any empirical evidence in opposition to this property can be viewed as an anomaly with respect to the baseline model.  
One of the workhorses to reveal anomalies is \textit{sorting}, whereby assets are ranked according to some particular attribute and portfolios are built based on these ranks (low values versus high values of the attribute).\footnote{Anomalies can also be revealed by regressions and we refer to \cite{baker2017detecting} for a detailed account on this matter.} If, over a sufficiently long period, the average returns of the sorted portfolios are statistically different, it is assumed that an anomaly is uncovered (see, e.g., \cite{cattaneo2020characteristic} for a theoretical account on statistical tests). Often, tests are performed via the null hypothesis that a long-short portfolio has a zero return, with for instance stocks with high values clustered in the long leg minus stocks with low values aggregated in the short leg. In our framework, this equally-weighted portfolio has a return equal to
\begin{align}
{r}_{t+1,LS}^{(k)}&= r_{t+1,+}-r_{t+1,-}= N_+^{-1}\sum_{n_+=1}^{N_+}r_{t+1,n_+}  -  N_-^{-1}\sum_{n_-=1}^{N_-}r_{t+1,n_-}   , \label{eq:rtk}
\end{align}
where $n_+$ and $n_-$ are the indices of the assets with high versus low value of a particular characteristic.\footnote{Strictly speaking, the cross-section aggregation of \textit{logarithmic} returns is not rigorous. Nevertheless, we proceed with this approximation. One case where it would be too coarse is if the sorting characteristic is the volatility of returns. } Without loss of generality, let us consider that the sought anomaly relates to the $k^{th}$ characteristic, which explains the superscript in the l.h.s. of Equation \eqref{eq:rtk}. For simplicity, we also assume that $N_+=N_-=N/2$ so that each leg of the portfolio consists of half of the investment universe. This is a simple and common choice (see, e.g., \cite{barberis2003style}), though other fractions, such as thirds, quintiles and deciles are popular alternatives in the asset pricing literature. We then get, using the first identity of the lemma,
\begin{align*}
\footnotesize
{r}_{t+1,LS}^{(k)}&=\frac{2}{N}   \sum_{n_\pm}\left[ \sum_{k=1}^{K} \left(\beta_{t+1}^{(k)}(c_{t,n_+}^{(k)}-c_{t,n_-}^{(k)}) + \eta_t^{(k)}(\Delta c_{t,n_+}^{(k)}-\Delta c_{t,n_-}^{(k)}) \right) \right]+\Lambda^{(k)}_{t+1}+ \Xi^{(k)}_{t+1},  \\
&=   \sum_{k=1}^{K}\left( \beta_{t+1}^{(k)} \Psi_t^{(k)} + \eta_t^{(k)} \Phi_t^{(k)}\right)+\Lambda^{(k)}_{t+1}+ \Xi^{(k)}_{t+1},
\end{align*}
with $\Psi_t^{(k)} =\frac{2}{N}\left(\sum_{n_+}c_{t,n_+}^{(k)}-\sum_{n_-}c_{t,n_-}^{(k)}\right)$ being the net portfolio average of characteristics $k$ and $ \Phi_t^{(k)}=\frac{2}{N}\left(\sum_{n_+}\Delta c_{t,n_+}^{(k)}-\sum_{n_-}\Delta c_{t,n_-}^{(k)} \right)$ the corresponding change thereof. The last terms in the expressions $\Lambda_{t+1}^{(k)}=\frac{2}{N}\left(\sum_{n_+}\alpha_{t+1,n_+} -\sum_{n_-}\alpha_{t+1,n_-}\right)$ and $\Xi_{t+1}^{(k)}=\frac{2}{N}\left(\sum_{n_+}\varepsilon_{t+1,n_+}-\sum_{n_-}\varepsilon_{t+1,n_-}\right)$ average the non-characteristic related demands and the pure supply-side components of the portfolio respectively, and, for simplicity, we assume that 
\begin{equation}
\E\left[\Xi_{t+1}^{(k)}\right]=0,
\label{eq:AE}
\end{equation}
so that the unconditional average of time-$t$ returns is
\begin{equation}
\bar{r}_{t+1,LS}^{(k)}=\mathbb{E}\left[{r}_{t,LS}\right]=\mathbb{E} \left[\Lambda_{t+1}^{(k)}+\underbrace{ \beta_{t+1}^{(k)} \Psi_t^{(k)} + \eta_t^{(k)} \Phi_t^{(k)}}_{\text{component driven by char. }k} + \underbrace{\sum_{j\neq k}\beta_{t+1}^{(j)} \Psi_t^{(j)} + \eta_t^{(j)} \Phi_t^{(j)}}_{\text{components driven by other chars.}}\right].
\label{eq:rbar0}
\end{equation}

The hypothesis that  the shifts in the supply side are null on average in Equation \eqref{eq:AE} is again a byproduct of our focusing solely on the demand side. The sources of anomalies that are not driven by characteristics come from $\mathbb{E} \left[\Lambda^{(k)}_{t+1}\right]$ only, which depends on $k$ only through the sorting procedure. In addition, from an estimation standpoint, if we allow for fixed effects in a panel model, then the average error per asset will be zero, which implies that $\Xi_{t+1}^{(k)}=0$ pointwise, that is, for each characteristic $k$ and estimation sample.

We then write $\mu_X^{(k)}=\E[X_t^{(j)}]$, for the means of random variables, where $X=\{\Psi, \Phi, \beta, \eta \}$, and $\sigma_{X,Y}^{(k)}=\E[(X_t^{(k)}-\mu_t^{(k)})(Y_s^{(j)}-\mu_s^{(j)})]$ for the covariance terms, where chronological indices remain flexible to adapt for possible time shifts (e.g., for $\beta$ and $\Psi$). The identity 
\begin{equation}
\E[XY]= \sigma_{X,Y}+\mu_X\mu_Y
\label{eq:id}
\end{equation}
implies
\begin{equation}
\bar{r}_{t+1,LS}^{(k)}=\mathbb{E}\left[\Lambda_{t+1}^{(k)}\right] +\sigma^{(k)}_{\beta,\Psi}+\mu^{(k)}_\beta\mu^{(k)}_\Psi+\sigma^{(k)}_{\eta,\Phi}+\mu^{(k)}_\eta\mu^{(k)}_\Phi+\sum_{j\neq k} \left( \sigma^{(j)}_{\beta,\Psi}+\mu^{(j)}_\beta\mu^{(j)}_\Psi+\sigma^{(j)}_{\eta,\Phi}+\mu^{(j)}_\eta\mu^{(j)}_\Phi\right) .
\label{eq:r2}
\end{equation}


\subsection{Discussion on distributions}


In order to gain further intuition, we must discuss some distributional properties of the elements in Equation \eqref{eq:rbar0}. The first major assumption that we make is  
\begin{equation}
\mu^{(j)}_\Phi = \mathbb{E}\left[\Phi_t^{(j)}\right]=0 , 
\label{eq:zeromean}
\end{equation}
because there is no reason why, a priori, a sorted portfolio should see a non-zero shift in the variation of its characteristic scores. This will be confirmed by the data subsequently.

The second important hypothesis we make is on the distribution of characteristics. We suppose that, at for any given time, the data has been processed such that, across the cross-section of stocks, characteristic $k$ has a standard Gaussian distribution (with zero mean and unit variance). In recent articles in asset pricing (\cite{kelly2019characteristics}, \cite{freyberger2020dissecting}), characteristics have uniform distributions. From that, it is easy to recover Gaussian laws by applying the inverse cumulative distribution function to the properly ``\textit{uniformized}'' values of the characteristics. More generally, we posit that the cross-section of characteristics follows a multivariate Gaussian law:
\begin{equation}
\bm{c}_t \overset{d}{=} \mathcal{N}(\bm{0}, \bm{\Sigma}_t), \quad \text{with} \quad \left[ \bm{\Sigma}_t \right]_{i,i}=1  \text{ and }  \left[ \bm{\Sigma}_t \right]_{j,k}=\rho_t^{(j,k)}.
\end{equation}

If the number of assets, $N$, is large enough, we can then approximate the aggregate portfolio score for characteristic $j=\{1,\dots,K\}$ with
\begin{equation}
\small
\Psi_t^{(j)} =\frac{2}{N}\left(\sum_{n_+}c_{t,n_+}^{(j)}-\sum_{n_-}c_{t,n_-}^{(j)}\right) \underset{N \rightarrow \infty}{\longrightarrow} \mathbb{E}_{\rho_t^{(j,k)}}\left[c_t^{(j)}|c_t^{(k)}>0\right]- \mathbb{E}_{\rho_t^{(j,k)}}\left[c_t^{(j)}|c_t^{(k)}<0\right]=\frac{4}{\sqrt{2\pi}}\rho_t^{(j,k)},
\label{eq:psi}
\end{equation}
where the zero threshold in the conditional expectations is the median value of the characteristic (we recall that the sorting procedure operates on characteristic $k$). Under Gaussian assumptions, the strong law of large numbers warrants that the left part converges almost surely to $\rho_t^{(j,k)}$, which is the time-$t$ correlation between characteristic $j$ and characteristic $k$ (see Appendix \ref{sec:idexp} for a proof). One key point is that the left-hand side of the above equation is random, thus we are obliged to envisage that the correlation term is stochastic as well. Empirically, this makes perfectly sense because correlations between characteristics are likely to vary with time.\footnote{In machine learning, the change in the covariance structure of predictors is referred to as \textit{covariate shift} - see, e.g., \cite{moreno2012unifying}.} Technically, this is why, in Equation \eqref{eq:psi}, the expectations are conditional on this correlation value.

 This is however not true for $\Psi_t^{(k)}$, which remains constant because when $j=k$ there is no interaction with other characteristics and the aggregate portfolio score is solely driven by the way the characteristics are normalized. Thus, $\Psi_t^{(k)} =4(2\pi)^{-1/2}$ is fixed and equal to its mean. 

A similar reasoning can be applied to $\Phi_t^{(j)} $, which captures the aggregate change in characteristic score of the long-short portfolio. Again, we assume that these changes are normally distributed, with zero mean, but we do not impose any covariance structure. The overarching distribution between the vectors $\Phi_t^{(j)}$ and $\Psi_t^{(j)}$ is a $2K$ Gaussian vector for which some rows and columns of the covariance matrix are filled with zeros. Indeed, by definition, almost surely, $\Phi_t^{(j)}=0$ for $j = k$. 

In turn, when $N$ is large enough,
$$\Phi_t^{(j)} =\frac{2}{N}\left(\sum_{n_+}\Delta c_{t,n_+}^{(j)}-\sum_{n_-}\Delta c_{t,n_-}^{(j)}\right) \underset{N \rightarrow \infty}{\longrightarrow} \frac{4}{\sqrt{2\pi}}\rho_{t,\Delta}^{(j,k)}\mathbb{V}\left[\Delta c_{t}^{(j)}\right]^{1/2},$$
where $\rho^{\Delta} _{j,k}$ is the correlation between the \textit{change} $\Delta c_{t,n}^{(j)}$ and the \textit{level} of the characteristic $ c_{t,n}^{(k)}$. The scaling constant comes from the identity in Appendix \ref{sec:idexp}.
The bilinearity of the covariance operator implies the following result, which shows that the average return of a characteristic-sorted portfolio depends on the relationships with all other characteristics.

\begin{lemma}
\label{lem:3}
Assume that \eqref{eq:AE} and \eqref{eq:zeromean} hold and that characteristics and their variations follow centered Gaussian distributions. As the number of assets increases to infinity, we have that
\begin{equation}
\footnotesize
\bar{r}_{t+1,LS}^{(k)} \underset{N\rightarrow \infty}{\longrightarrow}  \mathbb{E}\left[\Lambda_{t+1}^{(k)}\right]+ \frac{4}{\sqrt{2\pi}}  \left(\mu^{(k)}_\beta + \mathbb{C}\textnormal{ov}\left(\sum_{j\neq k}\beta_{t+1}^{(j)},\sum_{j\neq k}\rho_t^{(j,k)} \right)+ \mathbb{C}\textnormal{ov}\left(\sum_{j\neq k}\eta_t^{(j)},\sum_{j\neq k}\rho_{t,\Delta}^{(j,k)} \mathbb{V}\left[\Delta c_{t}^{(j)}\right]^{1/2}\right) \right) .
\label{eq:r3}
\end{equation}
\end{lemma}
The lemma suggests that the average return is captured by four components:
\begin{enumerate}
\setlength{\itemsep}{-2pt}
\item A term that depends on the characteristic $k$ independently from the demand in this characteristic (and in any characteristic).
\item The average demand for the characteristic.
\item The covariance between \textit{i)} summed demand shifts for all other characteristics and \textit{ii)} the summed correlation between characteristic $k$ and all other characteristics;
\item  The covariance between \textit{i)} total past (scaled) demands for all other characteristics and \textit{ii)} the summed (over $j$) scaled correlations between shifts in characteristic $j$ and the level of characteristic $k$.
\end{enumerate}

If characteristics are demeaned, $\E[XY]+\E[XZ]=\E[X(Y+Z)]$ implies that the third component is positive when the aggregate change in demand for all other characteristics is positively correlated with the correlation between the characteristic and the sum of all other characteristics. The fourth term is positive when the demand in all other characteristics is positively related to the scaled correlation between the characteristic and the summed changes in all other characteristics.

\section{Estimation}

\label{sec:esti}

The fundamental goal of the paper is to propose simple decompositions of asset returns. Given the linear forms obtained above, we opt for simple panel models which have intuitive outputs and interpretations. A notebook with the code used to generate our results and a link to the dataset is available \href{http://www.gcoqueret.com/files/misc/demand_6_share.html}{here}.


\subsection{Data}

The data originates from a major financial data provider and encompasses stock traded in the US. We prefer to share anonymized data, rather than be transparent about the source without being able to distribute the material. It is plain that data from another vendor, encompassing a different cross-section of assets, would generate alternative estimates. The purpose here is not to draw definitive conclusions about which characteristics drive returns, but to illustrate decompositions and approximate relative demands.
  
Time ranges between October 1994 and February 2021 and observations are sampled at a monthly frequency. There are 3,430 firms in total, with a minimum of 853 at the end of 1994 and a maximum of 2,878 in 2016. The sample includes 634,140 month-firm pairs. To ease readability when displaying our results, we focus on three firm characteristics: firm size, price-to-book ratio and past performance (prior 12 month to 1 month return) as a proxy for momentum. To evaluate the sensitivity of estimates to the set of characteristics, we will subsequently include three additional indicators: realized volatility over the past month, asset growth over the past year, and profitability margin. They correspond, respectively, to the low risk, investment, and profitability anomalies. The latter two were reported in \cite{fama2015five}.

First, these characteristics are normalized so that, for each month, they follow a standard Gaussian law (winsorized beyond $\pm3$). This simply requires to apply the empirical distribution function to the raw values, and to then proceed with the quantile function of the Gaussian law. Then, we evaluate differences in characteristics to obtain the $\Delta c_{t,n}^{(k)}$. For the sake of consistency with the convention of the paper, logarithmic returns are used instead of arithmetic ones.

\subsection{Results for panel models}

The linear relationships outlined in Lemma \ref{lem:1} call for an estimation of scaled demands and changes in demands - though the former have a superior interest from an asset pricing standpoint. Returns, characteristics (scores) and changes in characteristics are observed so that the $\beta_{t}^{(k)}$ and $\eta_t^{(k)}$ are the unknowns. The nature of the data imply a panel approach to estimation. 

\begin{figure}[!h]
\begin{center}
\includegraphics[width=15.1cm]{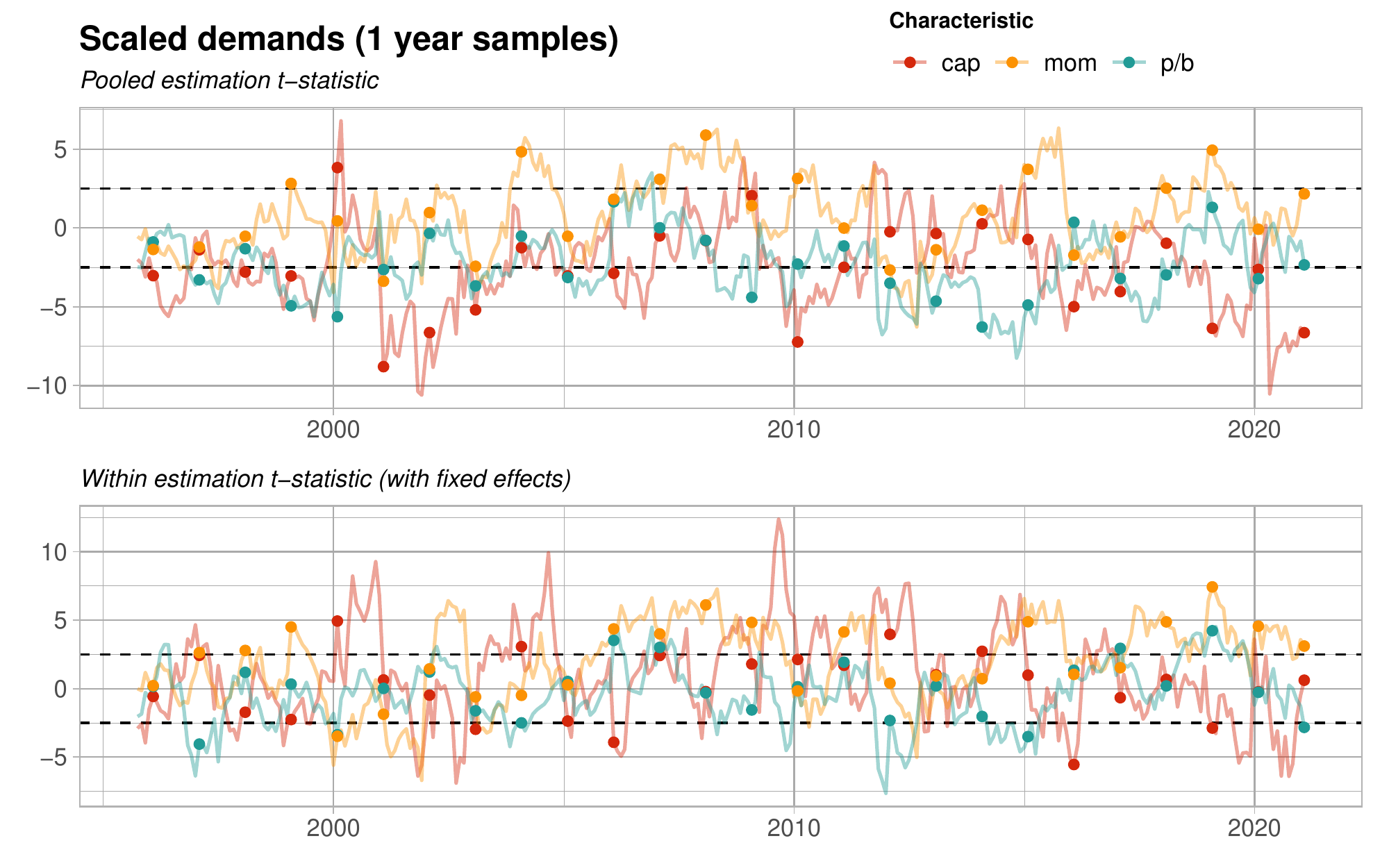}\vspace{-3mm}
\includegraphics[width=15.1cm]{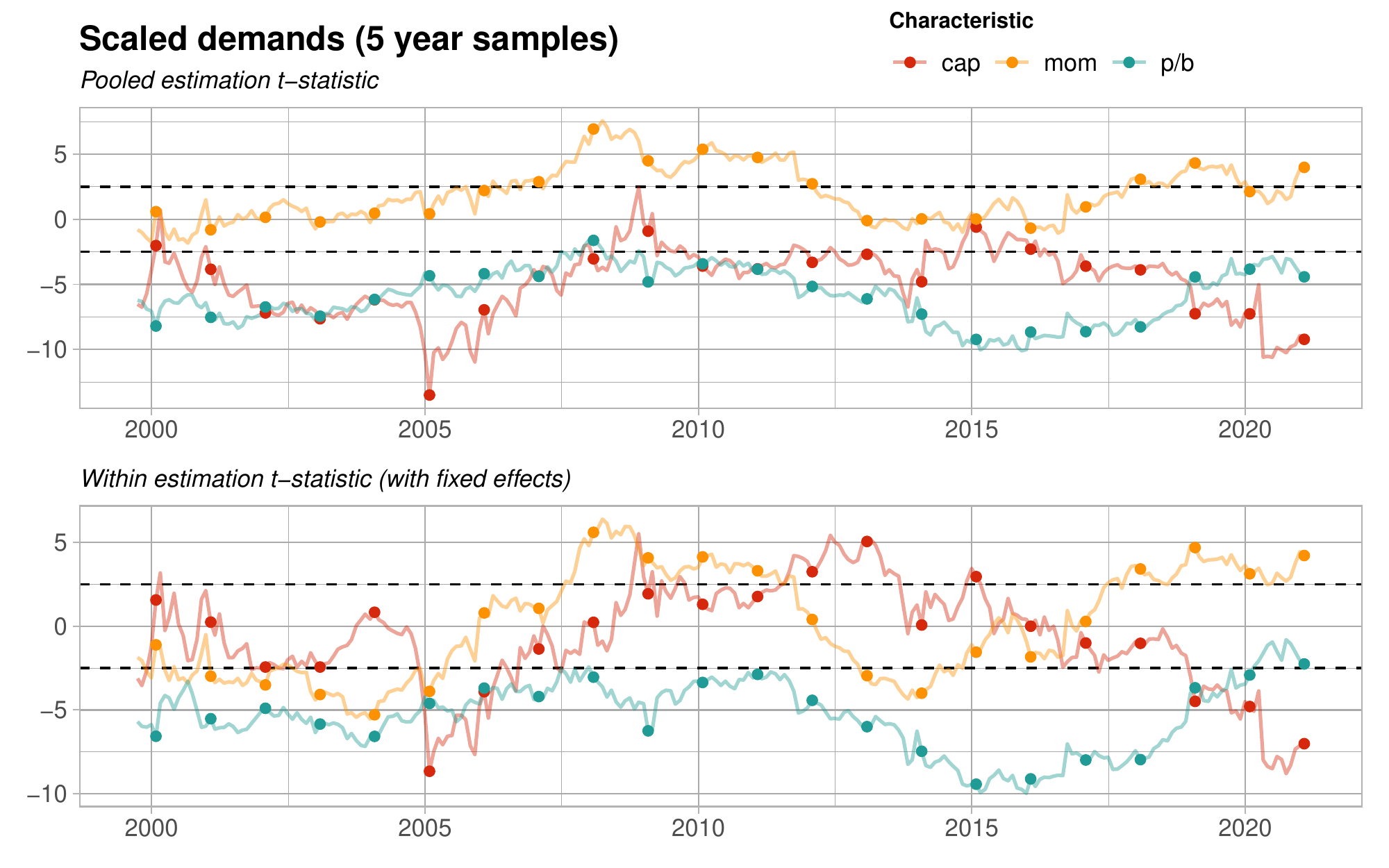}\vspace{-6mm}
\caption{\small{\textbf{Estimates for demands}. We plot the time-series of $\hat{\eta}_{t}^{(k)}$ estimated from Equation \eqref{eq:simple} using two methods: the \textbf{pooled} version ($\alpha_{t+1,n}=\alpha$) in the upper panels and the \textbf{within} version (with fixed effects: $\alpha_{t+1,n}=\alpha_n$) in the lower panel. In the upper graphs, estimations are performed on rolling samples of 12 months  so that 2 consecutive values are built on samples with an 11 month overlap. Each value for the month of January is shown with a small circle so that two consecutive circles relate to disjoint samples. In the lower graphs, estimations are run on 60 months and overlaps occur between four consecutive circles. Dashed black lines mark the $\pm2.5$ thresholds, which correspond to a 1\% significance level.} \label{fig:dem}}
\end{center}
\end{figure}

In Figure \ref{fig:dem}, we plot the $t$-statistics pertaining to the $\hat{\eta}_{t}^{(k)}$ estimates of Equation \eqref{eq:simple}. We resort to two alternative estimation methods: simple pooling (upper subpanels) and fixed effects (lower subpanels). The latter resorts to the least squares dummy variables (LSDV) estimator. We do not consider random effect models because the stock-specific constant in the expression of the returns is a quantity that should be estimated, just like the loadings on the other variables.  

In our study, we use rolling annual estimations samples in the upper panel and five year samples in the lower panel. 
The values when considering Equation \eqref{eq:simpl2} as estimation baseline are qualitatively similar. We do not produce them as they do not provide incremental added value. We plot the $t$-statistics instead of coefficient values because they are a better indicator of the significance of the demands for characteristics. The level of agreement between the two panels is arguably time-varying.

\begin{figure}[!h]
\begin{center}
\includegraphics[width=16cm]{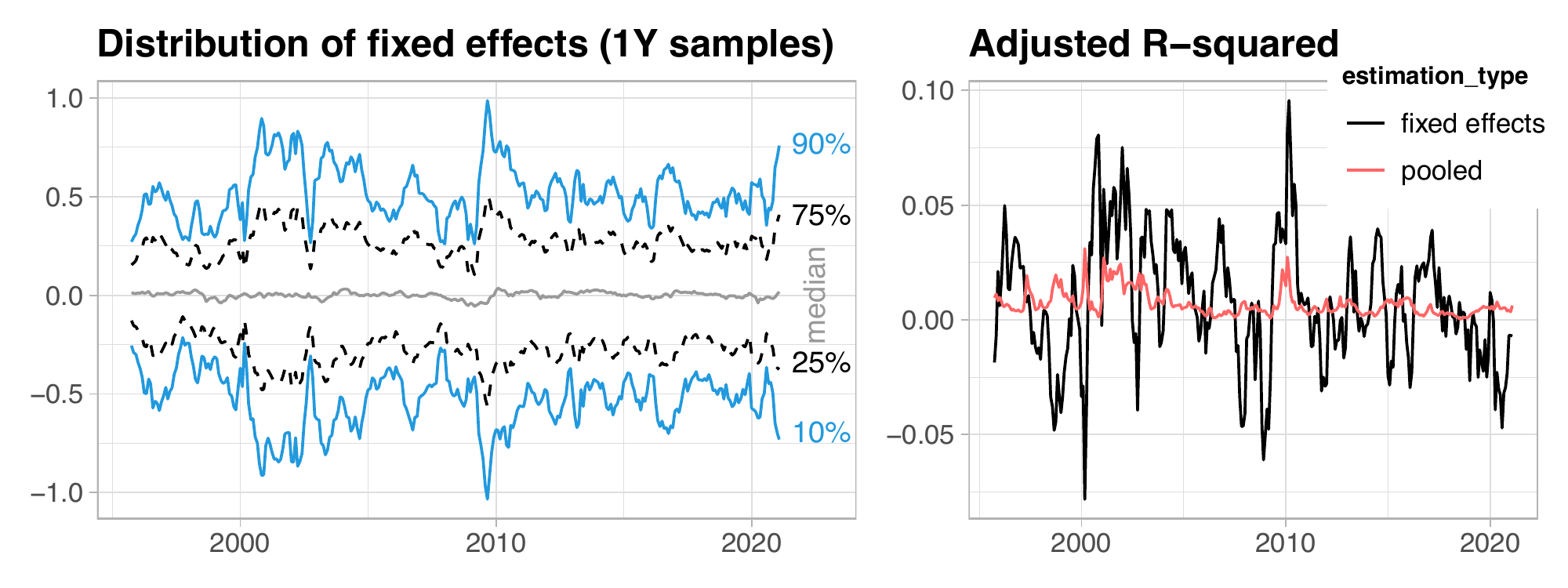}\vspace{-1mm}
\includegraphics[width=16cm]{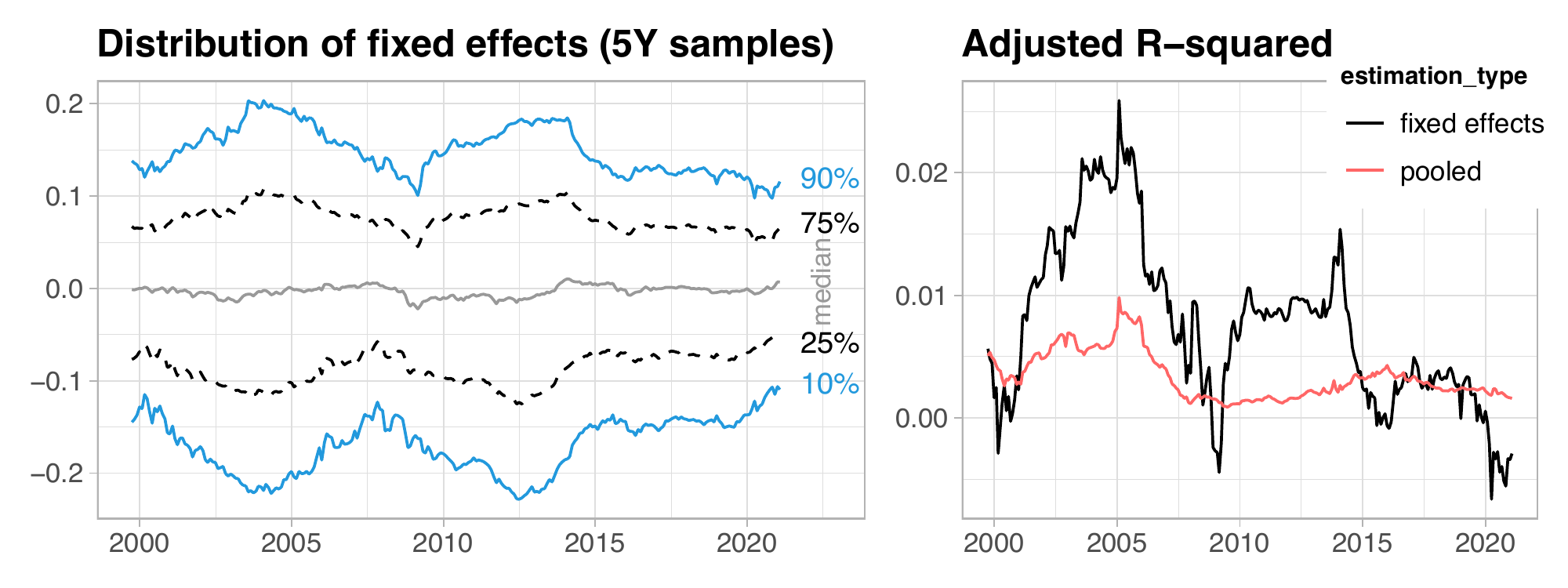}\vspace{-5mm}
\caption{\small{\textbf{Fixed effects and $R^2$}. We plot the time-series of cross-sectional quantiles of $\hat{\alpha}_{t,n}$ estimated from Equation \eqref{eq:simple} using fixed effects. In the first top (\textit{resp.}, bottom) panel, estimations are performed on rolling samples of 12 months (\textit{resp.}, 60 months) so that 2 consecutive values are built on samples with an 11 month overlap (\textit{resp.}, 59 month overlap). } \label{fig:fe}}
\end{center}
\end{figure}

To shed some light on this matter, we produce in Figure \ref{fig:fe} the time-series of some quantiles of the estimated fixed effects ($\hat{\alpha}_{t,n}$ - estimated from Equation \eqref{eq:simple}). It seems that the dispersion in fixed effects partly explains the degree to which the two subpanels in Figure \ref{fig:dem} yield similar results. The highest dispersion in $\hat{\alpha}_{t,n}$ occurs at the end of 2009, which is a point where the two estimations methods tend to disagree: the pooled estimation for the demand in cap is the highest, while for the within estimates are very low. In 2015, when the dispersion of fixed effects is low, both types of estimates tend to tell the same story. More fundamentally, the dispersion of fixed effects is an important indicator because it reveals if the independent variables are able to capture and explain the diversity in the cross-section of returns. 

Overall, on the long samples, the price-to-book ratio has a significant negative demand. The coefficients for market capitalization fluctuate but are also often negative. The estimates for momentum oscillate around zero, except for the pooled method on the long samples: in this case, most coefficients are positive.

The series in the lower panel are smoother by construction (because of longer sample overlaps) and reflect longer trends. For both estimation methods, the price-to-book and capitalization characteristic are predominantly associated with negative values. This indicates that investors prefer small, underpriced, stocks. For momentum, the two subpanels disagree: the pooled values are mostly positive, while the within estimates fluctuate around zero. 

In Figure \ref{fig:demadd} in Appendix \ref{sec:add}, we plot the same estimates when we include three additional characteristics (and their changes). For pooled estimates, it is hard to tell the difference when comparing to Figure \ref{fig:dem}. The differences exist, but are too marginal to be distinguished. For within $t$-statistics, the impact is substantial, which highlights the importance of the set of characteristics. When more of them are included, the estimated coefficients are smaller in magnitude. 

Finally, in unreported results, we estimated synchronous models in which the time indices of characteristics is the same as those of the returns. The outcome is qualitatively similar to our baseline findings. The only major change is the demand in the price-to-book, which is less negative with the long samples in the fixed effect models. These results are available upon request and can be reproduced with the online material.

\subsection{Estimation issues for anomaly decomposition}

Lemma \ref{lem:3} decomposes the average return of long-short portfolios based on characteristic sorting. There are four terms: the idiosyncratic demand term (based on fixed effects posterior to estimation), one average demand term and two covariance terms. Each one of them requires the computation of an expectation. In order to evaluate them, we must obtain time-series of the quantities inside the expectation operators in the lemma. 
A significant problem that we encounter is that the magnitude of some estimates is highly dependent on the sample size that is chosen to perform the rolling estimates.  

If we stack the coefficients into $\bm{\zeta}=(\bm{\beta},\bm{\eta})$, then Equation (10.50) in \cite{wooldridge2010econometric} implies that 
\begin{equation}
\hat{\bm{\zeta}}=\left(\sum_{n=1}^N\sum_{t=1}^T\ddot{\bm{z}}_{t,n}'\ddot{\bm{z}}_{t,n}\right)^{-1}\left( \sum_{n=1}^N\sum_{t=1}^T\ddot{\bm{z}}_{t,n}'\ddot{r}_{t+1,n}\right) ,
\label{eq:lsdv}
\end{equation}
where $\ddot{\bm{z}}_{t,n}$ is the vector of predictors that is demeaned (column-wise) at the stock level and $\ddot{r}_{t+1,n}$ is the similarly demeaned future return. 

According to Equation (10.58) in \cite{wooldridge2010econometric}, the expression for fixed-effects obtained from the LSDV estimator is 
\begin{equation}
\hat{\alpha}_{t+1,n}  = \bar{r}_{t+1,n}- \sum_{k=1}^{K} \left(\hat{\beta}_{t+1}^{(k)}\bar{c}_{t,n}^{(k)} + \hat{\eta}_t^{(k)}\bar{\Delta c}_{t,n}^{(k)} \right), 
\label{eq:fixef}
\end{equation}
where $\hat{\beta}_{t+1}^{(k)}$ and $\hat{\eta}_t^{(k)}$ are the LSDV estimates from \eqref{eq:lsdv} and $\bar{r}_{t+1,n}$, $\bar{c}_{t,n}^{(k)}$ and $\bar{\Delta c}_{t,n}^{(k)} $ the sample average of returns, characteristics, and their changes.

In Figure \ref{fig:fe}, the magnitude of fixed effect for small sample sizes (upper left panel) is several orders larger than that pertaining to longer sample sizes (lower left panel). This sizeable dispersion in fixed effects has at least two origins. First, the estimated coefficients. When samples are small, the inverse in \eqref{eq:lsdv} is not well conditioned, which implies larger magnitudes for $\hat{\bm{\zeta}}$ (i.e., $\hat{\bm{\beta}}$ and $\hat{\bm{\eta}}$), all other things equal. Indeed, small sample sizes imply that the smallest eigenvalues of the matrix to be inverted are close to zero. Upon inversion, the related values become arbitrarily large, which generates large elements in the matrix and thus in the estimates.\footnote{This is a very qualitative statement. Let us assume that the predictors are i.i.d. and multivariate Gaussian. While asymptotic results are well covered by the literature when the samples are large, the law of the spectrum of covariance matrices is less documented for finite samples (see \cite{rudelson2010non}). The distribution of the smallest eigenvalue (which matters substantially in the conditioning of the matrix) is treated in \cite{edelman1991distribution}. Further results on  condition numbers are obtained in \cite{edelman1988eigenvalues} and \cite{edelman2005tails}. }

The second source of dispersion lies in the sample averages in \eqref{eq:fixef}. Under an i.i.d. assumption on data generation, it is easy to show that large sample sizes decrease the variance of cross-sectional average. In Appendix \ref{app:disp}, we formally prove this statement and show that multiplying the sample size five-fold (from 12 to 60) divides the dispersion (standard deviation) by two on average. This creates a strong dependence on the sample size for the estimation of $\Lambda_{t+1}^{(k)}$ in Equation \eqref{eq:r3}, which we depict in Figure \ref{fig:fe3} in Appendix \ref{app:disp}.

\begin{figure}[!h]
\begin{center}
\includegraphics[width=16cm]{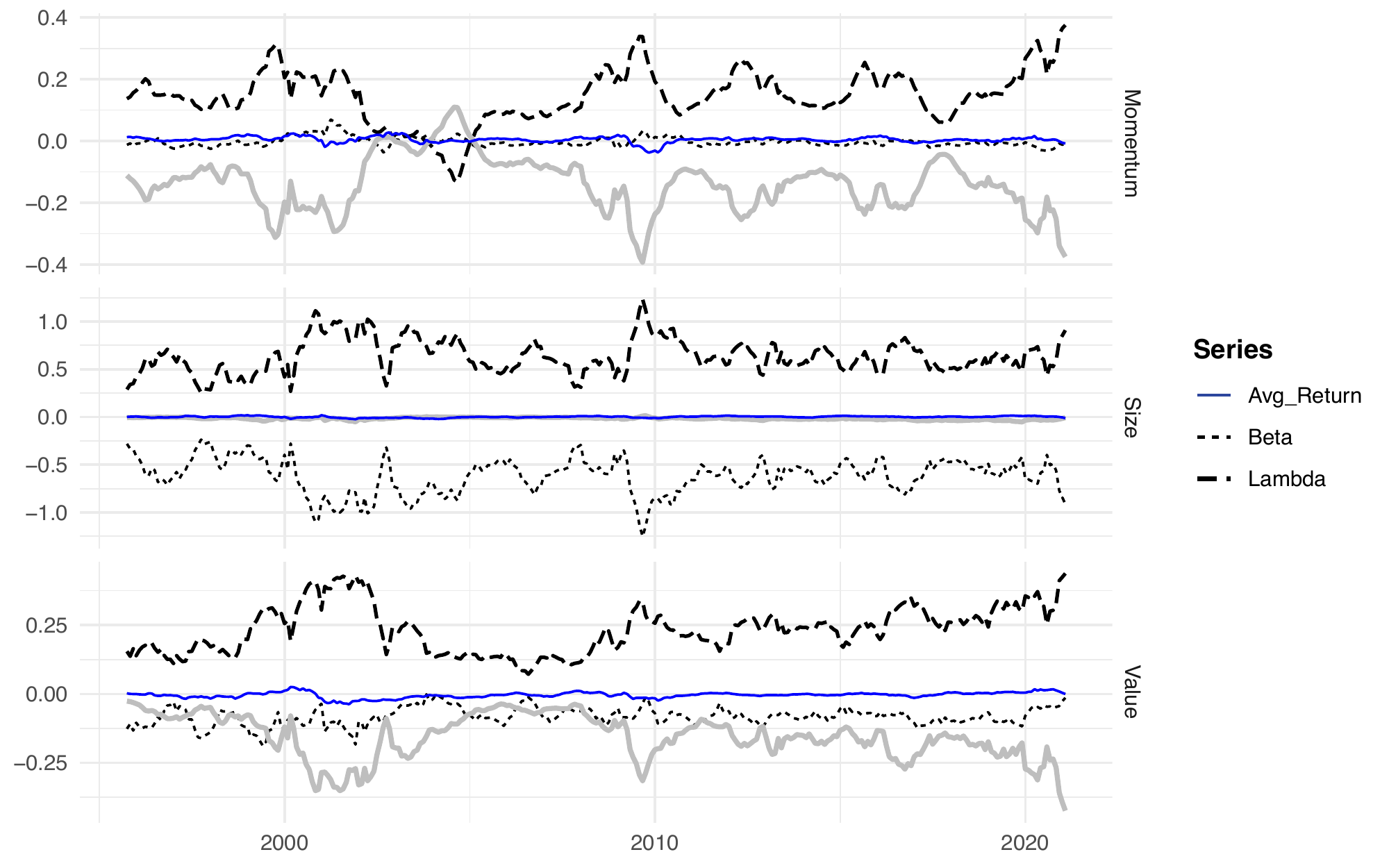}\vspace{-4mm}
\caption{\small{\textbf{Anomaly decomposition for one year samples}. We plot the time-series of the sample average of $\bar{r}_{t+1,LS}^{(k)}$ (thin blue line), of sample estimates of $\hat{\beta}_{t+1}^{(k)}$ (times $4/\sqrt{\pi}$) and $\hat{\Lambda}_{t+1}^{(k)}$ (dotted lines). The thick grey curves show the average return minus the two terms (loadings and fixed effects), and approximate the sum of covariance terms in \eqref{eq:r3}.  Estimates are computed for each sample, but portfolio compositions are adjusted monthly. } \label{fig:decompo}}
\end{center}
\end{figure}

As an illustration of these issues, we plot estimates of $\Lambda_{t+1}^{(k)}$ and $\beta_{t+1}^{(k)}$ in Figure \ref{fig:decompo}, along with average returns of rolling samples. The grey lines show the difference between the average returns and the sum of the two terms and thus serve as proxy for the last two terms in \eqref{eq:r3}, which consist of covariances. For the size factor (middle panel), the loadings and fixed effects cancel out, meaning that the covariance terms are possibly marginal. For the other two anomalies, this is not the case and fixed effects outweigh the loadings in amplitude, leaving room for the covariance interactions. The results are qualitatively the same for five year samples, but the magnitudes of fixed effects are approximately divided by five. Overall, the compensation of strong the fixed effects requires important interaction terms for the value and momentum anomalies.

The influential scale of fixed effects is a likely consequence of our resorting to a limited number of predictors. A handful\footnote{The dispersion of fixed effects is not reduced by switching from three to six predictors.} of characteristics are not enough to explain the cross-section of returns and other forces are at play that are not captured by the set of variables we have used in this study. This resonates with the empirical conclusions of \cite{koijen2019demand}, who also note that latent demands is the number one driver of the cross-section of returns.\footnote{They write: ``\textit{changes in latent demand are the most important,
explaining 81 percent of the cross-sectional variance of stock returns}.} Whether a broader set of characteristics would tame the importance of fixed effects is left for future work.

\subsection{Nonlinear demands: a discussion}

If we go back to the general formulation of returns from Equation \eqref{eq:center}, we can simplify the expression to
\begin{equation}
r_{t+1,n}=a^\Delta_{t+1,n}+ \sum_{i=1}^I B_{t+1,i} g_{t+1,i}(\bm{c}_{t,n})-\sum_{i=1}^I B_{t,i}g_{t,i}(\bm{c}_{t-1,n})+ \epsilon_{t+1,n},
\label{eq:nn}
\end{equation}
where $a^\Delta_{t+1,n}=\sum_{i=1}^I (B_{t+1,i} a_{t+1,i,n}-B_{t,i}a_{t,i,n})$ is the stock-specific, characteristics-independent, constant. Econometrically, the above equation could be interpreted for example:
\begin{enumerate}
\setlength{\itemsep}{-2pt}
\item as two ensembles of each $I$ models (e.g., random forests with $I$ trees)\footnote{An interesting question would be whether or not to allow for column/predictor subsampling during training. This would allow investors to form their portfolios based on an incomplete set of characteristics, which is a realistic assumption. } or
\item as two neural networks, each with $I$ units in their last layer (before aggregation) and each unit taking as input a possibly complex sub-network to generate nonlinearities (in $g_{t+1}$ and $g_t$).\footnote{Combinations of neural networks have recently emerged in the asset pricing literature, see for example \cite{gu2021autoencoder} in a different context.} 
\end{enumerate}

From an estimation standpoint, the second interpretation is easier to implement. The joint training of two separate tree models based on their own predictors is a complex task with little, if any, documentation. In contrast, the versatility of neural networks allows to train a sophisticated architecture like the one mentioned above. In any case, the global model requires to fix the number of agents, $I$. This is arguably an important degree of freedom that is likely to impact the estimates. 

With fixed estimation samples, and given the universal approximation property of neural networks, it is possible to increase the number of investors $I$ until the errors, become marginal. Whether the constants $a^\Delta_{t+1,n}$ would also fade remains an open question. They are the biases in the last layer of the overarching network. Nevertheless, the complexity of networks will preclude any simple interpretation like the ones obtained from the linear models studied in the present article. Tools from the explainable AI toolkit (e.g., feature importance or partial dependence plots, see \cite{molnar2020interpretable}) would become valuable to conclude on the sign and relative magnitude of individual factors. This is left for future research.

\clearpage

\section{Conclusion}

\label{sec:conc}

In this article, we derive several similar expressions for asset returns that are linear in assets' characteristics and changes thereof. The main modelling assumption is that agents have preferences that can be expressed in two separable components: the logarithm of the price of the asset, and a linear combination of its characteristics. 

From an estimation standpoint, returns and characteristics are given, and loadings are estimated via panel regressions with fixed effects. The only novelty compared to standard models is that we include \textit{changes} in characteristics. According to the equilibrium formula, the coefficients pertaining to these changes are the scaled demands for characteristics. 

Our contribution is theoretical in nature and proposes an interpretation for estimates of panel models in asset pricing. Our empirical results rely on rolling estimates that crucially depend on sample sizes. Given the amplitude of fixed effects in all of our models, our only robust conclusion is that a handful of characteristics is not enough to efficiently explain the cross-section of stock returns, at least with linear models.

We see two promising directions for future empirical research. The first is the inclusion of nonlinearities in demands, which would imply similar nonlinearities in returns. This is advocated by \cite{kirby2020firm} and tested successfully for predictive purposes in \cite{gu2020empirical}. However, the mainstream supervised algorithms used in the latter article are not sufficient to decompose demands in the sense of the present paper. 

Finally, the number of characteristics is a plausible route towards improved fit. The natural extension is to simply increase the set of predictors, but a model that handles the \textit{dynamic} inclusion of new characteristics would make more sense. The rise of alternative data based on sustainability, sentiment, now-casted earnings, etc., paves the way to sophisticated models in which the number of features slowly increases with time - as technology provides investors with myriads of characteristics.

\clearpage

\bibliographystyle{chicago}
\bibliography{bib}

\clearpage

\appendix
\section{Identity on a conditional expectation}
\label{sec:idexp}
Consider a bivariate Gaussian distribution $(Y,Z)$ with zero mean, variances $\sigma_y^2$ and $\sigma_z^2$, and correlation $\rho\in [-1,1]$. $Y$ is the sorting variable. We are interested in 
$$\frac{\E[Z 1_{\{Y >m\}}]}{P[Y>m]}-\frac{\E[Z1_{\{Y<m \}}]}{P[Y<m]}.$$
If $m$ is the median of $Y$ (here, zero), this simplifies to 
\begin{align*}
2\E[Z(1_{\{Y>0\}}-1_{\{Y<0 \}})] &= 2\E[Z(1_{\{Y>0\}}-1_{\{Y<0 \}})] =2\E[Z(1-2\times 1_{\{Y<0 \}})] = 4\E[Z1_{\{Y>0\}}] \\
&= 4\int_0^\infty \frac{e^{-y^2/(2\sigma_y^2(1-\rho^2))}}{2\pi \sigma_y^2\sigma_z^2 \sqrt{1-\rho^2}} \left(\int_\R ze^{-(z^2-2\rho yz\sigma_z/\sigma_y)/(2\sigma_z^2(1-\rho^2))}dz\right) dy \\
&= 4\int_0^\infty \frac{e^{-y^2/(2\sigma_y^2(1-\rho^2))}}{2\pi \sigma_y\sigma_z \sqrt{1-\rho^2}} \left(\rho y\frac{\sigma_z}{\sigma_y} \sqrt{2\sigma_z^2\pi(1-\rho^2)}e^{(\rho y)^2/(2\sigma_y^2\sigma_z^2(1-\rho^2))} \right)dy \\
&= 4\rho \frac{\sigma_z}{\sigma_y^2} \int_0^\infty y \frac{e^{-y^2/(2\sigma_y^2)}}{\sqrt{2 \pi}} dy = \frac{4}{\sqrt{2\pi}}\rho \sigma_z
\end{align*}

where the integral result in the third line comes from \cite{gradshteyn2007table}, equation 3.462-6, and the final equality is relatively standard and it is a simple case of equation 3.462-5.

\section{Linear demands}
\label{sec:lindem}

Agent $i$ believes the returns are driven by: $\bm{r}_{t+1}=\bm{C}_t\bm{\beta}_{t+1,i}+ \bm{e}_{t+1}$, where the $(N\times 1)$ vector of errors is independent from all other terms and has a zero mean vector and a covariance matrix $\text{diag}(\bm{\sigma}_{e,i}^2)$, where $\bm{\sigma}^2_{e,i}$ is the vector of variances of errors. Thus,
\begin{align}
\bar{\bm{r}}_t=\mathbb{E}_{t,i}[\bm{r}_{t+1}]&=\bm{C}_t\mathbb{E}_{t,i}[\bm{\beta}_{t+1,i}]=\bm{C}_t\hat{\bm{\beta}}_{t,i}, \label{eq:rt} \\
 \mathbb{V}_{t,i}[\bm{r}_{t+1}]&= \mathbb{E}_t[(\bm{r}_{t+1}-\bar{\bm{r}}_t)(\bm{r}_{t+1}-\bar{\bm{r}}_t)'] =\bm{C}_t \hat{\bm{\Sigma}}_{\bm{\beta},t,i}\bm{C}_t'- \bar{\bm{r}}_t\bar{\bm{r}}_t' + \text{diag}(\hat{\bm{\sigma}}_{e,i}^2)
\end{align}
with $ \hat{\bm{\Sigma}}_{\bm{\beta},t,i}=\mathbb{E}_{t,i}[\bm{\beta}_{t+1} \bm{\beta}_{t+1}']$ being agent $i$'s time-$t$ estimate covariance structure of the loadings. Likewise, $\hat{\bm{\beta}}_{t,i}$ is the time-$t$ estimate of the vector of expected loadings and $\hat{\bm{\sigma}}_{e,i}^2$ the expected (or estimated) variances of errors (we omit the time index for notational simplicity). \\
Using the Sherman-Morrison identity, we obtain a simplified expression for the estimation of the covariance matrix:
\begin{align}
 \mathbb{V}_{t,i}[\bm{r}_{t+1}]^{-1}= \bm{M}^{-1}\left( \bm{I}_N + \frac{\bar{\bm{r}}_t\bar{\bm{r}}_t'\bm{M}^{-1}}{1+\bar{\bm{r}}_t'\bm{M}^{-1}\bar{\bm{r}}_t}\right),
\end{align}
with $\bm{M}:=\bm{M}(\bm{C}_t,\hat{\bm{\Sigma}}_{\bm{\beta},t,i}, \hat{\bm{\sigma}}_{e,i}^2)=\bm{C}_t \hat{\bm{\Sigma}}_{\bm{\beta},t,i}\bm{C}_t'+ \text{diag}(\hat{\bm{\sigma}}_{e,i}^2)$. An application of a more general Woodbury identity\footnote{Namely, $(\bm{I}+\bm{AB})^{-1}=\bm{I}-\bm{A}(\bm{I}+\bm{BA})^{-1}\bm{B}$. Our application of this identity first factors out the diagonal matrix of error variance to obtain the identity matrix.} yields
\begin{equation}
\bm{M}^{-1}=\left(\bm{I}_N-\text{diag}(\hat{\bm{\sigma}}_{e,i}^2)^{-1} \bm{C}_t(\bm{I}_K+ \hat{\bm{\Sigma}}_{\bm{\beta},t,i}\bm{C}_t'\text{diag}(\hat{\bm{\sigma}}_{e,i}^2)^{-1}\bm{C}_t)^{-1} \hat{\bm{\Sigma}}_{\bm{\beta},t,i}\bm{C}_t' \right)\text{diag}(\hat{\bm{\sigma}}_{e,i}^2)^{-1}.
\label{eq:M}
\end{equation}
Now, the traditional mean-variance solution is
$$\bm{w}^*_{t,i}=\underset{\bm{w}}{\text{argmax}} \left\{\bm{w}'\bar{\bm{r}}_t - \frac{\gamma_{t,i}}{2}\bm{w}' \mathbb{V}_{t,i}[\bm{r}_{t+1}]\bm{w}, \text{ s.t } \bm{w}'\bm{1}=b\right\}=\gamma_{t,i}^{-1} \mathbb{V}_{t,i}[\bm{r}_{t+1}]^{-1}(\bar{\bm{r}}_t+ \delta_{t,i} \bm{1}),$$
where the constant $\delta_{t,i}$ is chosen to satisfy the budget constraint.
From this, we infer that the optimal weights satisfy the proportionality relationship
\begin{align}
\bm{w}^*_{t,i}(\bm{C}_t,\hat{\bm{\beta}}_{t,i},\hat{\bm{\Sigma}}_{\bm{\beta},t,i}, \hat{\bm{\sigma}}_{e,i}^2)& = \bm{M}^{-1}\left( \bm{I}_N + \frac{\bm{C}_t\hat{\bm{\beta}}_{t,i}\hat{\bm{\beta}}_{t,i}'\bm{C}_t'\bm{M}^{-1}}{1+\hat{\bm{\beta}}_{t,i}'\bm{C}_t'\bm{M}^{-1}\bm{C}_t\hat{\bm{\beta}}_{t,i}}\right)(\bm{C}_t\hat{\bm{\beta}}_{t,i}  +\delta_{t,i} \bm{1}).\nonumber
\end{align}
Thus, for one asset, the form of $\bm{M}^{-1}$ given in \eqref{eq:M} allows to write
$$w_{t,i,n} = f_{i,n,1}(\bm{C}_t,\hat{\bm{\beta}}_{t,i},\hat{\bm{\Sigma}}_{\bm{\beta},i}, \hat{\bm{\sigma}}_{e,i}^2)+\sum_{k=1}^Kc_{t,n}^{(k)}\times f_{i,n,2}(\bm{C}_t,\hat{\bm{\beta}}_{t,i},\hat{\bm{\Sigma}}_{\bm{\beta},i}, \hat{\bm{\sigma}}_{e,i}^2),$$
with 
\begin{align}
f_{i,n,2}& \propto -\hat{\sigma}_{e,i,n}^{-2} \left[ (\bm{I}_K+ \hat{\bm{\Sigma}}_{\bm{\beta},t,i}\bm{C}_t'\text{diag}(\hat{\bm{\sigma}}_{e,i}^2)^{-1}\bm{C}_t)^{-1} \hat{\bm{\Sigma}}_{\bm{\beta},t,i}\bm{C}_t' \text{diag}(\hat{\bm{\sigma}}_{e,i}^2)^{-1} \right. \\
& \hspace{21mm} \left. \times  \left( \bm{I}_N + \frac{\bm{C}_t\hat{\bm{\beta}}_{t,i}\hat{\bm{\beta}}_{t,i}'\bm{C}_t'\bm{M}^{-1}}{1+\hat{\bm{\beta}}_{t,i}'\bm{C}_t'\bm{M}^{-1}\bm{C}_t\hat{\bm{\beta}}_{t,i}}\right)(\bm{C}_t\hat{\bm{\beta}}_{t,i}+ \delta_{t,i}\bm{1} )\right]_{n,\cdot}   \\
f_{i,n,1}& \propto  -\hat{\sigma}_{e,i,n}^{-2}  \left[   \left( \bm{I}_N + \frac{\bm{C}_t\hat{\bm{\beta}}_{t,i}\hat{\bm{\beta}}_{t,i}'\bm{C}_t'\bm{M}^{-1}}{1+\hat{\bm{\beta}}_{t,i}'\bm{C}_t'\bm{M}^{-1}\bm{C}_t\hat{\bm{\beta}}_{t,i}}\right) \right]_{n,\cdot} (\bm{C}_t\hat{\bm{\beta}}_{t,i}  +\delta_{t,i}\bm{1})
\end{align}
where $\left[\bm{M} \right]_{n,\cdot}$ stands for the $n^{th}$ row vector of matrix $\bm{M}$ and $\hat{\sigma}_{e,i,n}$ is the $n^{th}$ element of $\hat{\bm{\sigma}}_{e,i}$.

\clearpage
\section{Dispersion of cross-sectional sample means}
\label{app:disp}

We consider a variable $z_{t,n}$ observed at time $t$ across $N$ dimensions indexed by $n$ (e.g., assets). We assume that the $N$-dimensional vector $\bm{z}_t$ is i.i.d. through time and follows some Gaussian distribution with mean vector $\bm{\mu}=(\mu_n)_{\{1 \le n\le N \}}$ and covariance matrix $\bm{\Omega}=(\omega_{ij})_{\{1 \le i,j\le N \}}$.
The average dispersion (variance) in cross-sectional sample means\footnote{When $T=1$, i.e., the dispersion is not over means but over simple returns, we refer to \cite{grant2016theoretical} for theoretical results.} is equal to
\begin{align*}
C &= \E\left[N^{-1} \sum_{n=1}^N \left( \sum_{t=1}^T\frac{z_{t,n}}{T}-\sum_{l=1}^N\sum_{t=1}^T\frac{z_{t,l}}{NT} \right)^2 \right] \\
&= \frac{1}{N}\sum_{n=1}^N \E \left[ \sum_{s,t} \frac{z_{t,n} z_{s,n}}{T^2}  + \sum_{l,m}\sum_{s,t}\frac{z_{t,l}z_{s,m}}{(NT)^2}  -2 \sum_{l=1}^N\sum_{s,t}\frac{z_{t,n}z_{s,l}}{NT^2}\right] \\
&=\frac{1}{N} \sum_{n=1}^N \left[\frac{\omega_{n,n}+T\mu_n^2}{T}+\sum_{l,m}\frac{\omega_{l,m}+T\mu_l\mu_m}{N^2T} -2 \sum_{l=1}^N\frac{\omega_{l,n}+T\mu_l \mu_n}{NT} \right] \\
&= \frac{1}{N} \sum_{n=1}^N \left[\frac{\omega_{n,n}+T\mu_n^2}{T} - \sum_{l=1}^N\frac{\omega_{l,n}+T\mu_l \mu_n}{NT} \right]  \\
&=\frac{1}{N}\sum_{n=1}^N\left( \mu_n^2- \frac{1}{N}\sum_{l=1}^N\mu_l\mu_n \right) + \frac{1}{NT}\sum_{n=1}^N\left(\omega_{n,n}- \frac{1}{N}\sum_{l=1}^N\omega_{l,n} \right)
\end{align*}
Both terms are positive for the same reasons. The second is, e.g., larger than $\frac{1}{2NT}\sum_{l,m}(\omega_{l,l}-\omega_{m,m})^2$.\footnote{This comes from the correlations being smaller than one in magnitude and, loosely speaking, from the identity $\sum_{i,j}(a_i^2-a_ia_j)=\frac{1}{2}\sum_{i,j}(a_i-a_j)^2$.} Therefore, as the sample size increases, the dispersion in sample means decreases. 

\begin{figure}[!h]
\begin{center}
\includegraphics[width=11cm]{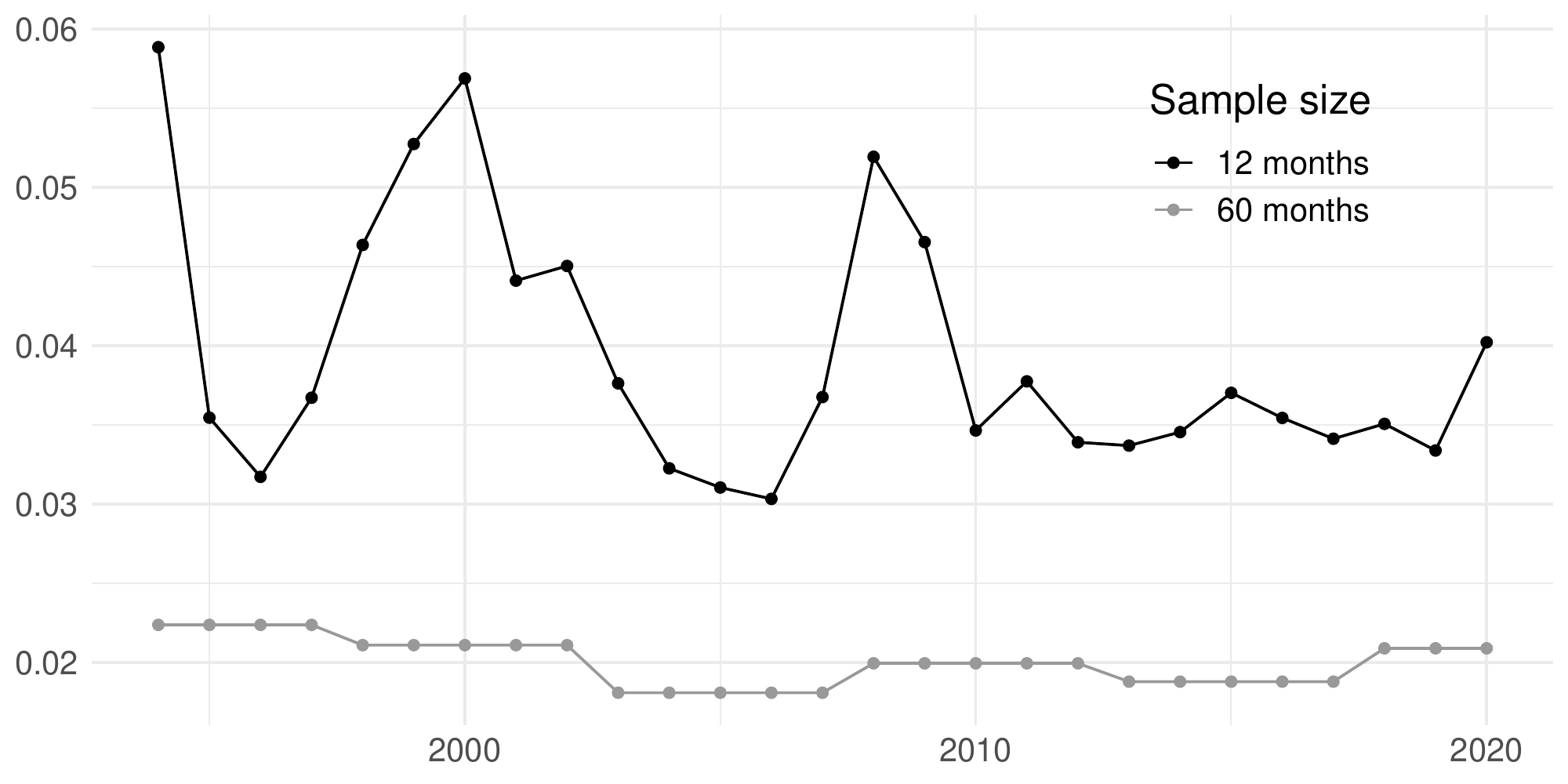} 
\vspace{-3mm}
\caption{\small{\textbf{Dispersion of sample means}. We plot the time-series of the standard deviation of average returns in the cross-section. Two sample sizes ($T$) are tested: 12 months and 60 months. The latter remain constant for five years. The values for 2021 are omitted because they are evaluated on an incomplete sample.} \label{fig:dispsample}}
\end{center}
\end{figure}

In Figure \ref{fig:dispsample}, we plot the time-series of the square root of the cross-sectional dispersion defined as  $N^{-1} \sum_{n=1}^N \left( \sum_{t=1}^T\frac{z_{t,n}}{T}-\sum_{l=1}^N\sum_{t=1}^T\frac{z_{t,l}}{NT} \right)^2$ when $z_{t,n}$ is the return of the stocks in the sample. We observe that multiplying the sample size by five divides the standard deviation by a factor two on average. This translates into a dispersion that is four times smaller for the deeper sample.

\begin{figure}[!h]
\begin{center}
\includegraphics[width=15cm]{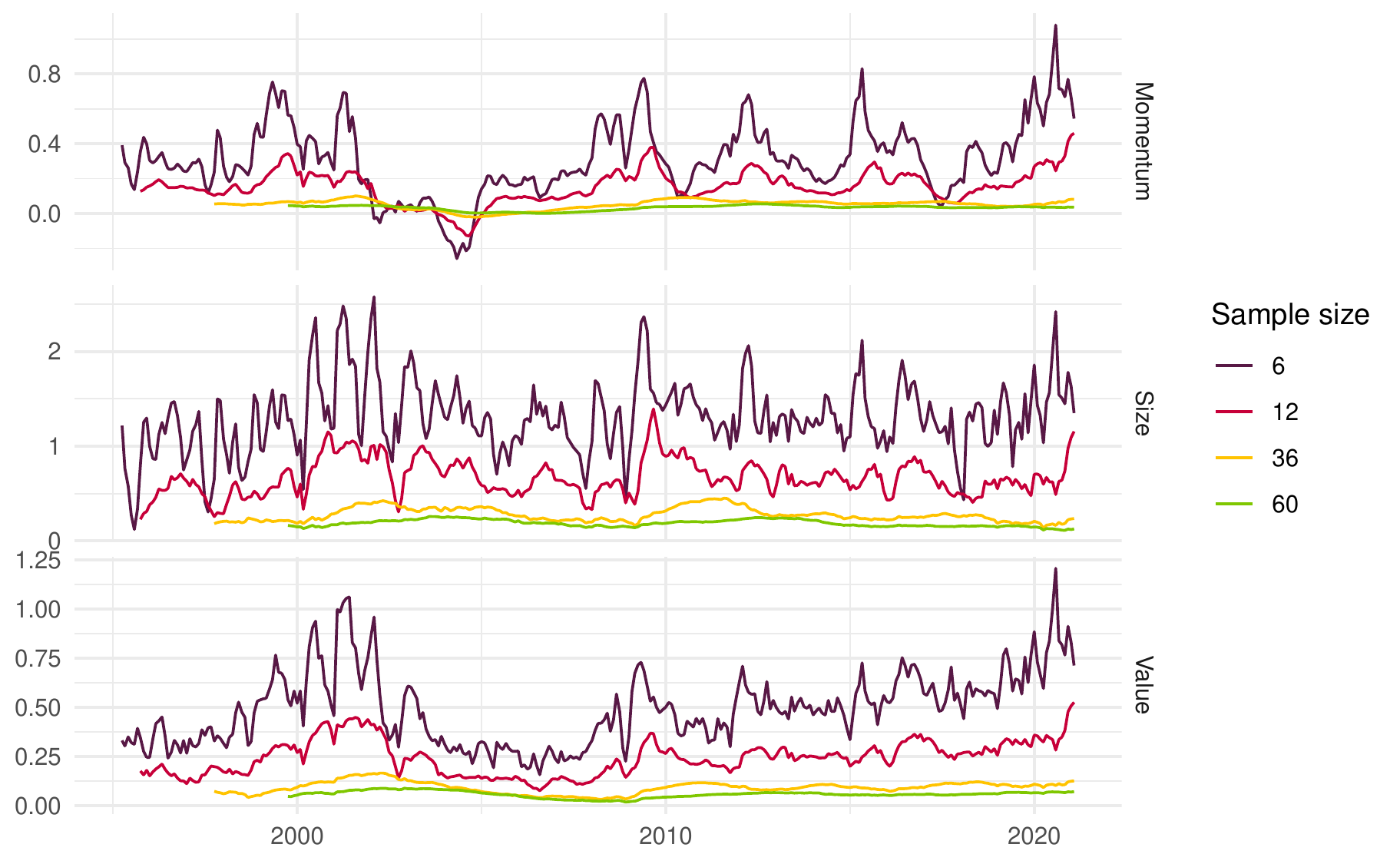}\vspace{-1mm}
\caption{\small{\textbf{Average fixed effects of sorted portfolios}. We plot the time-series of $\hat{\Lambda}_{t+1}^{(k)}$ in Equation \eqref{eq:r3}. For each point in time, we take a sample of the past $T$ observations and estimate the panel model with the three characteristics as well as their variations. Each portfolio consists of all of the sample and is split in halves, according to whether a stock is either small or large, or growth or value, or winner or loser.  We then take the average of fixed effects for the high leg of the factor/anomaly minus the average for the low leg (i.e., the portfolios are equally-weighted). Estimates are computed for each sample, but portfolio compositions are adjusted monthly. } \label{fig:fe3}}
\end{center}
\end{figure}

In Figure \ref{fig:fe3}, we plot the weighted average of fixed effects of portfolios sorted on market capitalization, price-to-book ratio and past returns. Plainly, the amplitude of the portfolios' mean fixed effect is decreasing with sample size. 

\clearpage

\section{Sensitivity to additional characteristics}
\label{sec:add}
\vspace{-3mm}

\begin{figure}[!h]
\begin{center}
\includegraphics[width=15.5cm]{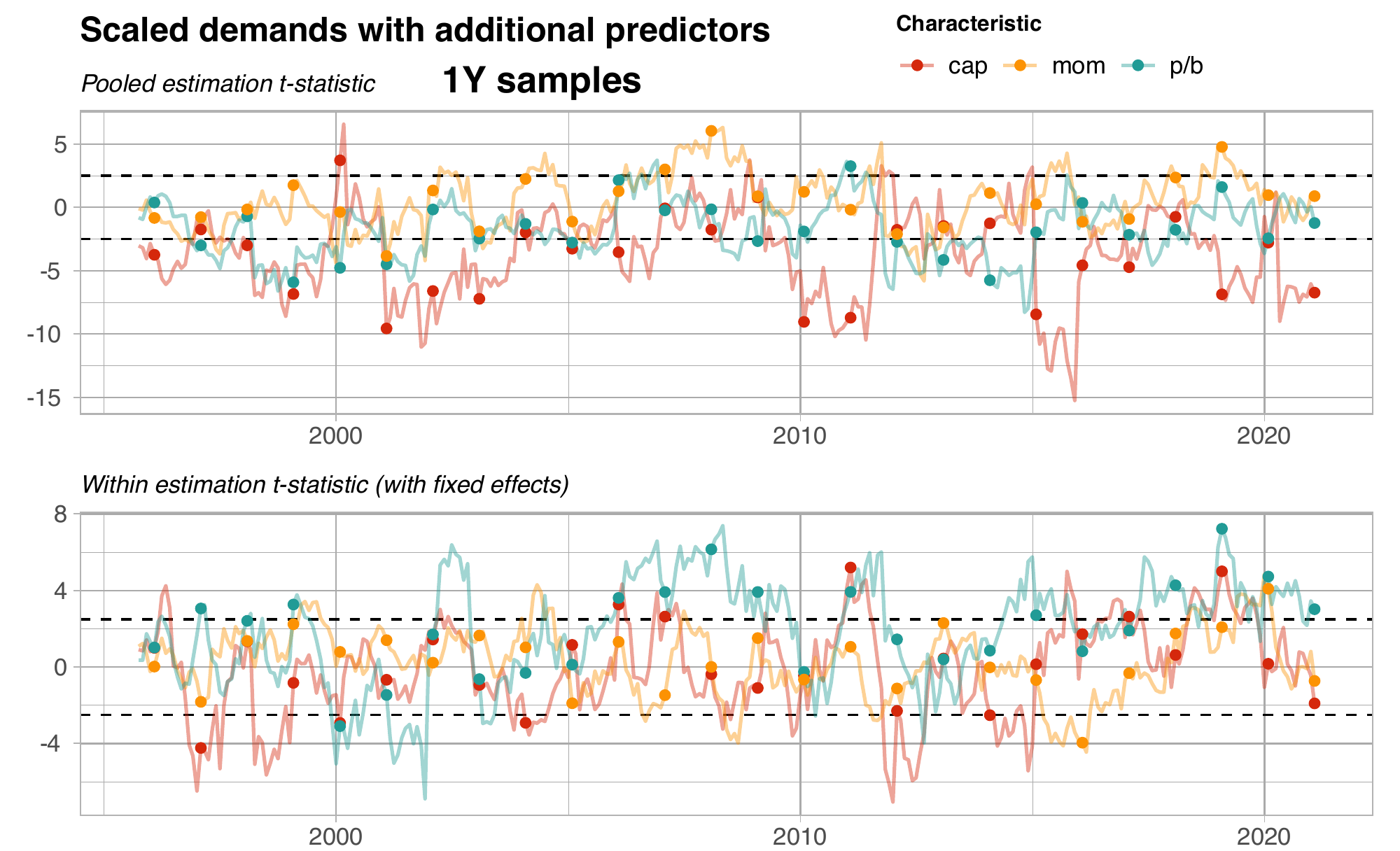} \vspace{-5mm}
\includegraphics[width=15.5cm]{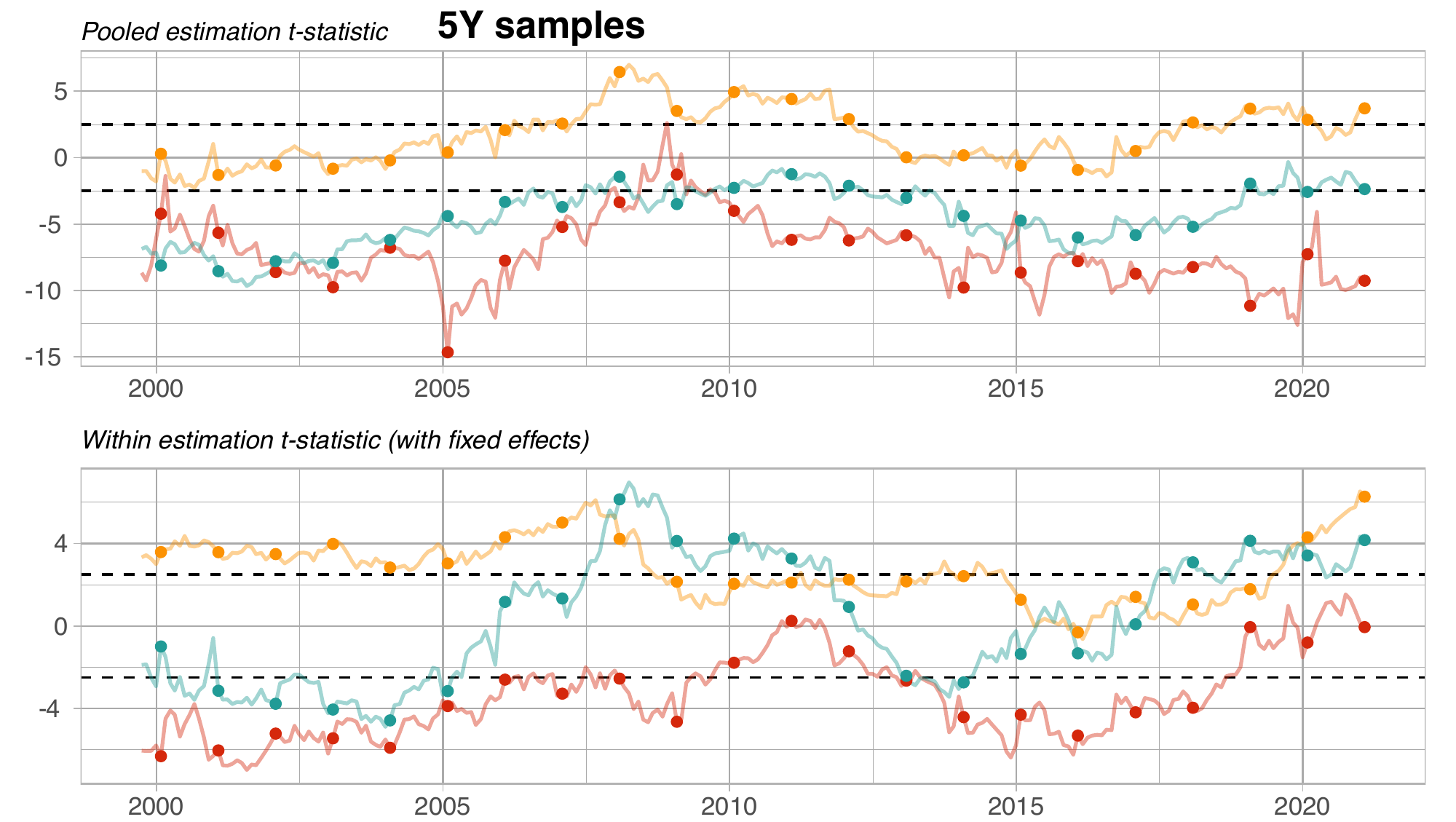}
\vspace{1mm}
\caption{\small{\textbf{Estimates for demands with additional characteristics}. We plot the time-series of $\hat{\eta}_{t}^{(k)}$ estimated from Equation \eqref{eq:simple}. Compared to Figure \ref{fig:dem}, three characteristics were added as predictors: profitability margin, asset growth, and realized volatility over the past month. Estimations are performed on rolling samples of 12 months (upper panel) and 60 months (lower panel). The time-series for the control characteristics can be viewed on the notebook of the paper.} \label{fig:demadd}}
\end{center}
\end{figure}

\end{document}